\documentclass[10pt,english,pre,nofootinbib,reprint,showpacs,showkeys,preprintnumbers,floatfix]{revtex4-1}
\usepackage[latin9]{inputenc}
\usepackage[a4paper]{geometry}
\usepackage{xcolor}
\usepackage{pdfcolmk}
\usepackage{babel}
\usepackage{units}
\usepackage{amsmath}
\usepackage{amssymb}
\usepackage{graphicx}
\usepackage{esint}
\PassOptionsToPackage{normalem}{ulem}
\usepackage{ulem}
\usepackage[unicode=true,
 bookmarks=true,bookmarksnumbered=false,bookmarksopen=false,
 breaklinks=false,pdfborder={0 0 1},backref=false,colorlinks=false]
 {hyperref}
\hypersetup{pdftitle={Information field dynamics for simulation scheme construction},
 pdfauthor={Torsten En{\ss}lin}}

\makeatletter

\providecolor{lyxadded}{rgb}{0,0,1}
\providecolor{lyxdeleted}{rgb}{1,0,0}
 \@ifundefined{textcolor}{}
 {%
   \definecolor{BLACK}{gray}{0}
   \definecolor{WHITE}{gray}{1}
   \definecolor{RED}{rgb}{1,0,0}
   \definecolor{GREEN}{rgb}{0,1,0}
   \definecolor{BLUE}{rgb}{0,0,1}
   \definecolor{CYAN}{cmyk}{1,0,0,0}
   \definecolor{MAGENTA}{cmyk}{0,1,0,0}
   \definecolor{YELLOW}{cmyk}{0,0,1,0}
 }
\newenvironment{lyxlist}[1]
{\begin{list}{}
{\settowidth{\labelwidth}{#1}
 \setlength{\leftmargin}{\labelwidth}
 \addtolength{\leftmargin}{\labelsep}
 }}
{\end{list}}

\usepackage{aas_macros}
\usepackage{bbold}

\newcommand{\g}{\mathcal{G}}
\newcommand{\p}{\mathcal{P}}
\newcommand{\D}{\mathcal{D}}

\makeatother

\begin{document}

\title{Information field dynamics for simulation scheme construction}

\author{Torsten A. Enßlin}

\email{ensslin@mpa-garching.mpg.de}

\selectlanguage{english}%

\affiliation{{\small Max Planck Institute for Astrophysics, Karl-Schwarzschildstr.1,
85741 Garching, Germany}}
\begin{abstract}
Information field dynamics (IFD) is introduced here as a framework
to derive numerical schemes for the simulation of physical and other
fields without assuming a particular sub-grid structure as many schemes
do. IFD constructs an ensemble of non-parametric sub-grid field configurations
from the combination of the data in computer memory, representing
constraints on possible field configurations, and prior assumptions
on the sub-grid field statistics. Each of these field configurations
can formally be evolved to a later moment since any differential operator
of the dynamics can act on fields living in continuous space. However,
these virtually evolved fields need again a representation by data
in computer memory. The maximum entropy principle of information theory
guides the construction of updated datasets via \textit{entropic matching},
optimally representing these field configurations at the later time.
The field dynamics thereby become represented by a finite set of evolution
equations for the data that can be solved numerically. The sub-grid
dynamics is thereby treated within auxiliary analytic considerations.
The resulting scheme acts solely on the data space. It should provide
a more accurate description of the physical field dynamics than simulation
schemes constructed ad-hoc, due to the more rigorous accounting of
sub-grid physics and the space discretization process. Assimilation
of measurement data into an IFD simulation is conceptually straightforward
since measurement and simulation data can just be merged. The IFD
approach is illustrated using the example of a coarsely discretized
representation of a thermally excited classical Klein-Gordon field.
This should pave the way towards the construction of schemes for more
complex systems like turbulent hydrodynamics. 
\end{abstract}



\maketitle

\section{Introduction}

\subsection{Motivation}

Computer simulations of fields play a major role in science, engineering,
economics, and many other areas of modern life. Computer limitations
require that the infinite number of degrees of freedom of a field
are represented by a finite data set that fits into computer memory.
For example in hydrodynamics with mesh codes, the average density,
pressure, and velocities of the fluid within grid cells form the data.
The data makes statements about the field properties, and the simulation
scheme describes how the present data determines the future data.
This dynamics is usually set up such that the continuum limit of an
infinite number of infinitesimal dense grid points recovers the partial
differential equations governing the physical field dynamics. However,
there are many possible schemes to discretize the differential operators
of the field equations. Which one gives good results already at finite
resolution? Which one takes the influence of processes on sub-grid
scales best into account? To address these questions, a rigorous approach
to construct accurate simulation schemes, information field dynamics
(IFD), is presented here. IFD rests on information field theory (IFT),
the theory of Bayesian inference on fields \citep{1999physics..12005L,2009PhRvD..80j5005E}.
In the ideal case, IFD and IFT provide identical results, since both
can be used to make statements about fields at later times given some
initial data. However, in real world applications of simulation schemes,
compromises with respect to accuracy and computational complexity
are often unavoidable. Thus IFD can be regarded as a particular approximation
scheme within IFT, which may or may not provide optimal results from
an information theoretical point of view.

The basic idea is that IFT turns the data in computer memory into
an ensemble of field configurations which are consistent with the
data and the knowledge on the sub-grid physics and field statistics.
The differential operators of the field dynamics can then formally
operate on these field configurations without the usual discretization
approximation. An unavoidable approximation finally happens when these
time evolved fields get recast into the finite data representation
in computer memory. The information theoretical guideline of the Maximum
Entropy Principle (MEP) is used in order to ensure maximal fidelity
of this operation, which we call in the following \textit{entropic
matching}. The sub-grid dynamics is thereby treated within an auxiliary
analytic consideration. In the end, an IFD simulation scheme for the
time evolution of a field is a pure data updating operation in computer
memory, and therefore an implementable algorithm. Although this algorithm
does not explicitly deal with a field living in continuous space any
more, it was, however, derived with the continuous space version of
the original problem being very present in the mathematical reasoning.
The sub-grid information, which IFT used to construct the virtual
continuous space field configurations, is encapsulated implicitly
in the resulting IFD scheme. Therefore IFD schemes act solely on the
data in computer memory without using any explicit sub-grid field
representation. 

When constructing a computational simulation scheme for field dynamics,
whether using IFD or not, one is facing two bottlenecks: finite computer
memory and finite computational time. This work deals only with the
first issue, and explains how to construct schemes which optimally
use the data stored in computer memory. Optimizing with respect to
only one objective, memory in this case, very often results in solutions
which are ineffective with respect to another aim, computational simplicity
here. Thus we do not expect the resulting IFD schemes necessarily
to be the optimal solution for a concrete computational problem. Deriving
practically usable schemes will often require additional approximations
in order to reduce the computational complexity. The IFD framework
can, however, help to clarify the nature of the approximations made
and guide the design of simulation schemes.

The concrete problem of how to discretize a thermally excited Klein-Gordon
(KG) field in position space will illustrate the usage of the theoretical
IFD framework.

\subsection{Previous work}

Our main motivation is to aid the construction of simulation schemes,
for example in hydrodynamics, for which a very rich body of previous
work exists. Appendix \ref{sec:Appendix----Previous} discusses briefly
the relevant concepts of partial differential equation discretization,
sub-grid modeling, and information theoretical concepts in simulation
schemes and their relation to IFD.

\subsection{Structure of this work}

In Sect. \ref{sec:Concepts} we introduce the necessary concepts of
IFT, MEP, and IFD. In Sect. \ref{sec:Information-field-dynamics}
IFD is developed in detail on an abstract level, as well as for the
illustrative example of a KG field. The fidelity of IFD and a typical
ad-hoc scheme for the KG field are compared numerically and against
an exact solution in Sect. \ref{sec:Numerical-verification}. Section
\ref{sec:Conclusions-and-outlook} contains our conclusion and outlook.

\section{Concepts\label{sec:Concepts}}

\subsection{Information field theory\label{sub:Information-field-theory}}

The idea of this work is that the data stored in a computer is only
a constraint on possible field configurations, but does not to fully
determine a unique sub-grid field configuration. Instead, the ensemble
of possible field configurations is constructed using IFT. IFT blends
the information in the data and any prior knowledge on the field behavior
into a single probability density function (PDF) over the space of
all field configurations.

IFT is information theory applied to fields, probabilistic reasoning
for an infinite set of unknowns, the field values at all space positions.
It provides field reconstructions from finite data. For this IFT needs
data, a data model describing how the data are determined by the field,
and a prior PDF summarizing the statistical knowledge on the field
degrees of freedom prior to the data. How this works in our case will
be shown in the following. A general introduction to IFT can be found
in \citep{2009PhRvD..80j5005E} and in the references therein.

IFT exploits mathematical methods from quantum and statistical field
theory. The unknown field $\phi$ is regarded as a signal, a hidden
message to be revealed from the data $d$. A prior PDF $\p(\phi)$
describes the knowledge about the signal field prior to the data,
and a likelihood PDF $\p(d|\phi)$ describes the probability of the
data given a specific signal field configuration. Bayes' theorem allows
one to construct the posterior PDF 
\begin{equation}
\p(\phi|d)=\frac{\p(d|\phi)\,\p(\phi)}{\p(d)},
\end{equation}
which summarizes the a posteriori (after the data is taken) knowledge
on the signal field. The connection to statistical field theory becomes
apparent, when one realizes that Bayes theorem can also be written
as 
\begin{equation}
\p(\phi|d)=\frac{e^{-H(d,\phi)}}{Z(d)},
\end{equation}
with the information Hamiltonian 
\begin{equation}
H(d,\phi)=-\log\p(d,\phi)=-\log\p(d|\phi)-\log\p(\phi),
\end{equation}
and the partition function 
\begin{equation}
Z(d)=\p(d)=\int\mathcal{D}\phi\,\p(d,\phi)=\int\mathcal{D}\phi\, e^{-H(d,\phi)}.
\end{equation}
Here, $\int\mathcal{D}\phi$ denotes a phase space integral over all
possible field configurations of $\phi$, a so called path integral. 

$ $The information Hamiltonian combines prior and likelihood into
a signal energy, which determines the signal posterior according to
the usual Boltzmann statistics. This Hamiltonian therefore contains
all available information on the signal field.

The simplest IFT case is that of a free theory. This emerges in case
three conditions are met: 
\begin{lyxlist}{00.00.0000}
\item [{(i)}] The a priori distribution of the field is a multivariate
Gaussian, 
\begin{equation}
\p(\phi)=\mathcal{G}(\phi,\Phi)=\frac{1}{\sqrt{|2\pi\Phi|}}\exp\left(-\frac{1}{2}\phi^{\dagger}\Phi^{-1}\phi\right),\label{eq:Gauss}
\end{equation}
with signal covariance $\Phi=\langle\phi\,\phi^{\dagger}\rangle_{(\phi)}=\int\mathcal{D}\phi\,\p(\phi)\,\phi\,\phi^{\dagger}$,
its determinant $|\Phi|=\mathrm{det}\Phi$, and $\phi^{\dagger}\psi=\int dx\:\overline{\phi}_{x}\,\psi_{x}$
denoting the scalar product. 
\item [{(ii)}] The data depends linearly on the signal field, 
\begin{equation}
d=R\,\phi+n,\label{eq:D=00003D00003D00003DRs+n}
\end{equation}
with a known response operator $R$. 
\item [{(iii)}] The noise $n=d-R\,\phi$ is signal-independent with Gaussian
statistics 
\begin{equation}
\p(n|\phi)=\mathcal{G}(n,N),
\end{equation}
where $N=\langle n\, n^{\dagger}\rangle_{(n|\phi)}=\int\mathcal{D}n\,\p(n|\phi)\, n\, n{}^{\dagger}$. 
\end{lyxlist}
In this case, the likelihood $\p(d|\phi)=\p(n=d-R\phi|\phi)=\g(d-R\phi,N)$
and the prior $\p(\phi)$ contribute terms to the Hamiltonian that
are at most quadratical in the signal. Thus, the Hamiltonian is also
quadratical, which is the mark of a free theory. In this specific
case, the information Hamiltonian states that the posterior field
is also Gaussian, but with shifted mean $m=\langle\phi\rangle_{(\phi|d)}=\int\mathcal{D}\phi\,\phi\,\mathcal{P}(\phi|d)$
and uncertainty variance $D$, which can be read off from 
\begin{eqnarray}
H(d,\phi) & \widehat{=} & \frac{1}{2}\left(d-R\phi\right)^{\dagger}N^{-1}\left(d-R\phi\right)+\frac{1}{2}\phi^{\dagger}\Phi^{-1}\phi^{\dagger}\nonumber \\
 & \widehat{=} & \frac{1}{2}[\phi^{\dagger}\underbrace{(\Phi{}^{-1}+R^{\dagger}N^{-1}R)}_{D^{-1}}\phi+\phi^{\dagger}\underbrace{R^{\dagger}N^{-1}d}_{j}+j^{\dagger}\phi]\nonumber \\
 & = & \frac{1}{2}\left(\phi^{\dagger}D^{-1}\phi+\phi^{\dagger}j+j^{\dagger}\phi\right)\nonumber \\
 & \widehat{=} & \frac{1}{2}\left(\phi-m\right)^{\dagger}D{}^{-1}\left(\phi-m\right),
\end{eqnarray}
with 
\begin{equation}
m=D\, j=\underbrace{\left(\Phi^{-1}+R^{\dagger}N^{-1}R\right)^{-1}R^{\dagger}N^{-1}}_{W}d=Wd.\label{eq:m=00003D00003D00003DWd}
\end{equation}
Here and later ``$\widehat{=}$'' means equality up to irrelevant
constants.%
\footnote{This is of course a context dependent convention, since it depends
on what is regarded to be relevant. In the context of this work, any
field dependent quantity is relevant. Field independent normalization
constants of PDFs are not. The sign ``$\widehat{=}$'' is here used
as the logarithmic partner of the sign ``$\propto$'', since normalization
constants become constant additive terms after taking the logarithms.
Later on, we will also regard terms of higher order in the time step
$\delta t$ as irrelevant, since they can be made to vanish by taking
the limit $\delta t\rightarrow0$. %
} In analogy to the quantum field theory, an information propagator
$D=(\Phi^{-1}+R^{\dagger}N^{-1}R)^{-1}$ and an information source
$j=R^{\dagger}N^{-1}d$ can be identified. The information source
$j$ is given by the data $d$, weighted by the inverse noise covariance
$N^{-1}$ and back-projected with the hermitian adjoint response $R^{\dagger}$
into the signal space. The a posteriori mean field $m_{x}$ at some
location $x$ of the signal space is constructed by transporting the
information $j_{y}$ sourced by the data at some location $y$ to
$x$ with the help of the information propagator $D_{xy}$. This happens
by applying this as a linear operator to the information source field
$m_{x}=\int dy\, D_{xy}\, j_{y}$. The resulting posterior mean field
depends linearly on the data, $m=Wd$. The corresponding linear filter
operation $W$ is well known in signal reconstruction as the (generalized)
Wiener filter \citep{1949wiener}. The information propagator $D$
is also identical to the a posteriori uncertainty variance, 
\begin{equation}
D=\langle(\phi-m)\,(\phi-m)^{\dagger}\rangle_{(\phi|d)},
\end{equation}
also known under the term Wiener variance. To conclude, in free IFT
the posterior is Gaussian with Wiener mean and variance, 
\begin{equation}
\p(\phi|d)=\mathcal{G}(\phi-m,D).
\end{equation}
Although the field mean $m$ is a continuous function in the signal
space, a full field with an apparently infinite number of field values,
it has strictly speaking only effectively a finite number of degrees
of freedom due to its construction. Since the mean field is a deterministic
function of the data, $m=m(d)=W\, d$, the phase space of possible
mean fields can have at most as many dimensions as the data has degrees
of freedom. This sets a limit to the maximal possible accuracy a simulation
scheme can achieve with finite data representation of the field. However,
in this work, we do not only evolve the mean field, but the full distribution
of plausible fields around this as characterized by $\mathcal{P}(\phi|d)$. 

It should be noted that there exist two equivalent formulations of
the Wiener filter operator 
\begin{eqnarray}
W & = & \left(\Phi^{-1}+R^{\dagger}N^{-1}R\right)^{-1}R^{\dagger}N^{-1}\nonumber \\
 & = & \Phi\, R^{\dagger}\left(R\,\Phi\, R^{\dagger}+N\right)^{-1}.
\end{eqnarray}
The first one is called the signal space and the second one the data
space representation, since the operator inversions happen in signal
and data space, respectively. They are fully equivalent as long as
$\Phi$ and $N$ are regular matrices.%
\footnote{The equivalence of the two Wiener filter representations is easily
verified via the following equivalence transformations: 
\begin{eqnarray*}
\left(\Phi^{-1}+R^{\dagger}N^{-1}R\right)^{-1}R^{\dagger}N^{-1} & = & \Phi\, R^{\dagger}\left(R\,\Phi\, R^{\dagger}+N\right)^{-1}\\
\Leftrightarrow R^{\dagger}N^{-1}\left(R\,\Phi\, R^{\dagger}+N\right) & = & \left(\Phi^{-1}+R^{\dagger}N^{-1}R\right)\Phi\, R^{\dagger}\\
\Leftrightarrow R^{\dagger}N^{-1}R\,\Phi\, R^{\dagger}+R^{\dagger} & = & R^{\dagger}+R^{\dagger}N^{-1}R\Phi\, R^{\dagger}.
\end{eqnarray*}
}

The data space representation of the Wiener filter $W=\Phi\, R\,\left(R\,\Phi\, R^{\dagger}+N\right)^{-1}$
can cope with the here relevant case of negligible noise, $N\rightarrow0$,
leading to $W=\Phi\, R\,\left(R\,\Phi\, R^{\dagger}\right)^{-1}$.
This is possible only if $\tilde{\Phi}=R\,\Phi\, R^{\dagger}$, the
data space image of the signal field covariance, is (pseudo)-invertible,
which is very often the case. If not, the data contains redundancies
that could be used to tailor the data space until $\tilde{\Phi}$
is invertible.

This noiseless limit might be a desirable assumption for dealing with
the data of a numerical simulation, since one might define the data
to represent a statement about the field like $d=R\,\phi$ exactly,
without any uncertainty in data space. However, in the course of a
field dynamical simulation, the knowledge of the exact field configuration
$\phi$ might not be present at later times due to unavoidable discretization
errors. Therefore, a mismatch of the data $d$ in computer memory
and the correct discretized statement $R\,\phi$ for the true field
might develop and this can be regarded as noise $n=d-R\,\phi$. Furthermore,
a full error propagation of initial value uncertainties in a simulation
might be of interest in case the initial data resulted from a real
measurement with instrumental noise. For these reasons, we will keep
the noise term in the formalism. 

The Wiener filter theory described so far gives us a sufficient IFT
background for this initial work on IFD. It should be noted, however,
that in case of non-linear relations between data and signal, or non-Gaussian
signal or noise statistics, IFT becomes an interacting field theory,
and the resulting operations on the data to calculate a posteriori
mean and variance become nonlinear. Such operations can be constructed
using diagrammatic perturbation series, re-summation and re-normalization
techniques \citep{2009PhRvD..80j5005E,2011PhRvD..83j5014E}, or by
the construction and minimization of an effective action, the Gibbs
free energy \citep{2010PhRvE..82e1112E,2011PhRvE..84d1118O}. In many
cases, the posterior is well approximated by a multivariate Gaussian,
which we assume in the following.

\subsection{Entropic matching}

We assume now that an ensemble of field configurations for a time
$t$ has been constructed with IFT, those being consistent with the
data $d=d_{t}$ and any background information at that time. It has
to be specified now how those evolve, and how this can be represented
by an updated dataset $d'=d_{t'}$ at a later time $t'$.

Each of the possible field configurations is assumed to evolve for
a short period according to the exact physical field dynamics. In
order to recast this evolved ensemble of field configurations back
into the data representation of the computational scheme, an updated
data set has to be constructed. The field ensemble implied by the
updated data should resemble the evolved field ensemble of the original
data as close as possible. We will use entropic matching for this,
the usage of the MEP without any additional constrains. The MEP is
the principle of our choice since it derives from very generic and
desirable first principles on how to update a probability without
introducing spurious knowledge.

For the MEP, entropy is just regarded as an abstract quantity that
can be used to rank various possible PDFs according to how well they
are suited to represent a knowledge state. A large entropy resembles
an uninformed or ignorance state. MEP aims therefore for the least
informed state that is still consistent with all known constraints.
This should be the state with the least spurious assumptions.

A number of intuitively obvious requirements on the internal logic
of such a ranking fully determines the functional form of this entropy
\citep{1957PhRv..108..171J,1957PhRv..106..620J,1982ieee...70..939J,2008arXiv0808.0012C}.
These requirements are that local information should have only local
effects, that the ranking should be independent of the coordinate
system used, and that independent systems lead to separable PDFs.
These requirements are further detailed in Appendix \ref{sec:Appendix----Maximum}.
The only function on the space of PDFs that is consistent with these
principles is the entropy 
\begin{equation}
\mathcal{S}(\p|\mathcal{Q})=-\int\mathcal{D}\phi\:\p(\phi)\:\log\left(\frac{\p(\phi)}{\mathcal{Q}(\phi)}\right),\label{eq:entropy}
\end{equation}
where $\p(\phi)$ denotes a PDF for some field $\phi$ to be ranked
for its ignorance, and $\mathcal{Q}(\phi)$ an a priori ignorance
state. This entropy is the relative entropy of information theory,
the Kullback-Leibler divergence of $\p$ to $\mathcal{Q}$ \citep{2008arXiv0808.0012C}.
It is in general also equivalent (up to some constant) to the Gibbs
energy of thermodynamics \citep{2010PhRvE..82e1112E}, and to the
Boltzmann-Shannon entropy in case the ignorance knowledge state $\mathcal{Q}$
does not favor any region of physical phase-space, i.e. $\mathcal{Q}(\phi)=\mathrm{const}$.

Since the information entropy is equivalent to the Kullback-Leibler
distance of information theory, it can also be used to match one PDF
optimally to another one. This entropic matching will be needed in
this work in order to find the data constrained representation of
the field PDF at a later instant that best matches the time evolved
PDF of an earlier instant. In case $\p(\phi)$ can be changed at any
phase-space point $\phi$, maximizing $\mathcal{S}(\p|\mathcal{Q})$
will reproduce the ignorance prior $\p\rightarrow\mathcal{Q}$. If
there are, however, constraints limiting the flexibility of $\p(\phi)$
to adapt to $\mathcal{Q}(\phi)$, the MEP solution will be different.
Such constraints can be imposed with the help of Lagrange multipliers,
respective thermodynamical potentials, which can be used to imprint
certain expectation values onto $\p$ as it is shown in Appendix \ref{sec:Appendix----Maximum}.
In this work, constraints arise due to the fact that the degrees of
freedom to represent functions and PDFs in computers are limited by
the size of the computer memory.

To be concrete, we write $\phi'=\phi_{t'}$ and assume for definiteness
only that the short time step $\delta t=t'-t$ permits a deterministic
and invertible functional relation between $\phi'$ and the earlier
$\phi=\phi_{t}$, so that $\p(\phi'|\phi)=\delta(\phi'-\phi'(\phi))$
as well as $\p(\phi|\phi')=\delta(\phi-\phi(\phi'))$. %
\footnote{Stochastic terms could easily be incorporated into the dynamics, e.g.
by setting $\p(\phi'|\phi)=\mathcal{G}(\phi'-\phi'(\phi),\,\delta t\,\Xi)$
in case of additive Gaussian and temporally white noise $\xi_{t}$
with covariance $\langle\xi_{t}\,\text{\ensuremath{\xi}}_{t'}^{\dagger}\rangle_{(\xi)}=\delta(t-t')\,\Xi$.
This is a straightforward extension of the scheme presented here \citep{2012arXiv1209.3700R}.%
}

Here and later, we assume further that the target knowledge state
$\mathcal{Q}$ in our case is given by the Gaussian signal field posterior
$\p(\phi|d,t)=\mathcal{G}(\phi-m,D)$ at time $t$ as specified by
the data $d=d_{t}$ and the background knowledge at this time, however
evolved according to the dynamical laws to a later time $t'$, so
that 
\begin{eqnarray}
\mathcal{Q}(\phi') & = & \p(\phi'|d)=\int\mathcal{D}\phi\,\p(\phi'|\phi)\,\p(\phi|d)\nonumber \\
 & = & \mathcal{G}(\phi(\phi')-m,D)\left|\frac{\partial\phi}{\partial\phi'}\right|.
\end{eqnarray}
The state $\p'$ we want to match to this using the MEP is one that
can be represented by a new set of data $d'=d_{t'}$ at this later
time via the IFT posterior $\p'(\phi')=\p(\phi'|d')=\mathcal{G}(\phi'-m',D')$.
Since the data degrees of freedom are finite, the PDF implied by this
new data (via $m'=W'd'$ and $D'=(\Phi'^{-1}+R'^{\dagger}N'^{-1}R')^{-1}$)
will be of a parametric form, with the new data being the parameters.
However, the evolved PDF will in general have a different functional
form. Therefore, a matching between the PDFs $\p'(\phi'|d')$ and
$\mathcal{Q}(\phi')$ is needed and using the MEP for this ensures
that the least amount of spurious information is introduced in this
unavoidable approximative step.

\subsection{Simulation schemes construction\label{sub:Simulation-schemes-construction}}

The IFD methodology to discretize the dynamics of a field can be summarized
with the following recipe: 
\begin{enumerate}
\item \textbf{Field dynamics:} The field dynamics equations have to be specified.
The KG equation, which can be derived from a suitable Hamiltonian,
will serve as an example in this work. 
\item \textbf{Prior knowledge:} The ignorance knowledge state in case of
the absence of data has to be specified. In our example the field
will be assumed to be initially excited by contact with a thermal
bath of known temperature. The Hamiltonian determining the field dynamics
will therefore also determine the background knowledge on the initial
state in our example. 
\item \textbf{Data constraints:} The relation of data and the ensemble of
field configurations being consistent with data and background knowledge
has to be established using IFT. Assimilation of external measurement
data into the simulation scheme is naturally done during this step. 
\item \textbf{Field evolution:} The evolution of the field ensemble over
a short time interval has to be described. This either involves the
evolution of the mean and spread of the ensemble, or \textemdash{}
as we will use here \textemdash{} the analytical description of the
evolution of all possible field configurations. 
\item \textbf{Prior update:} The background knowledge for the later time
has to be constructed. In the chosen example, energy and phase-space
conservation of the Hamiltonian dynamics guarantee that the same thermal
ignorance state also holds at later times. 
\item \textbf{Data update:} The relation of data and field ensemble has
to be invoked again to construct the data of the later time using
entropic matching based on the MEP. Thereby a transformation rule
is constructed that describes how the initial data determines the
later data. This transformation forms the desired numerical simulation
scheme. It has incorporated the physics of the sub-grid degrees of
freedom into operations solely in data space. 
\end{enumerate}
An IFD simulation scheme resulting from this recipe acts only on the
data space. Any sub-grid dynamics is encapsulated implicitly. This
is ensured by the auxiliary analytic considerations that construct
the ensemble of possible field configurations, evolve them analytically
in time, and map them back onto the data representation using entropic
matching.

\section{Information field dynamics\label{sec:Information-field-dynamics}}

The IFD program outlined above shall now be discussed in detail and
by following the recipe of Sect. \ref{sub:Simulation-schemes-construction}
step by step. The discussion will only deal with linear dynamics and
Gaussian knowledge states. Many interesting problems involve nonlinear
dynamics, and consequently should lead to non-Gaussian knowledge states.
However, the construction of a nonlinear IFD theory will have its
foundation in linear theory, which therefore needs to be developed
first.

$\blacktriangleright$ In order to illustrate the IFD methodology,
the problem of how to discretize the dynamics of a thermally excited
Klein-Gordon field in one-dimensional position space is chosen as
an example. Since exact solutions of the field dynamics can easily
be given in Fourier-space representation, an exact, sub-grid field
model exists in this case to which numerical solutions using IFD and
other discretization schemes can be compared. Passages dealing specifically
with this example are marked as this paragraph and might be skimmed
over on a first reading. $\blacktriangleleft$

\subsection{Field dynamics}

The linear dynamics of a field $\phi$ can in general be written as
\begin{equation}
\partial_{t}\phi=c+L\,\phi,\label{eq:linear dynamics}
\end{equation}
where $L$ is a linear operator acting on the field vector of a time
instance, thereby determining the field's time derivative. $L$ can
be a differential operator, it can include integro-differential operations,
and it can depend on time. A dependence on earlier field values is
excluded from $L,$ which is therefore assumed here to be local in
time. The field independent, but potentially time and position dependent
additive term $c$ is a source term of the field.

Nonlinear dynamics of the form 
\begin{equation}
\partial_{t}\chi=F(\chi)
\end{equation}
can often be cast approximatively into the form \eqref{eq:linear dynamics}
via a Fréchet-Taylor expansion around a sufficiently good and known
approximation $\psi$ for $\chi=\psi+\phi$: 
\begin{equation}
\partial_{t}\phi=\underbrace{F(\psi)-\partial_{t}\psi}_{c}+\underbrace{\partial_{\psi}F(\psi)}_{L}\,\phi+\mathcal{O}(\phi^{2}).
\end{equation}
One obvious choice of such an approximation would be to use a static
function $\psi_{t}=\chi_{t_{0}}$ for some short period $\left[t_{0},\, t_{1}\right]$
and afterwards $\psi_{t}=\chi_{t_{1}}$ for the next such period,
always ensuring $\phi$ to be small and second order effects to be
negligible.

Stochastic terms in the evolution equations can also be included into
the formalism, however, here we refrain from such complications and
assume fully deterministic dynamics. If higher time derivatives are
part of the linear or linearized evolution equation, these can be
included as further components of $\phi$.

$\blacktriangleright$ For example, the one dimensional Klein-Gordon
(KG) equation for a real scalar field with mass $\mu$ 
\begin{equation}
\partial_{t}^{2}\varphi=(\partial_{x}^{2}-\mu^{2})\varphi,
\end{equation}
which will serve as a concrete example in this work, can be cast into
the form \eqref{eq:linear dynamics} by setting $\phi=(\varphi^{\dagger},\pi^{\dagger})^{\dagger}$
and 
\begin{eqnarray}
\partial_{t}\begin{pmatrix}\varphi\\
\pi
\end{pmatrix} & = & L\,\phi=\begin{pmatrix}0 & 1\\
(\partial_{x}^{2}-\mu^{2}) & 0
\end{pmatrix}\begin{pmatrix}\varphi\\
\pi
\end{pmatrix}\nonumber \\
 & = & \begin{pmatrix}\pi\\
(\partial_{x}^{2}-\mu^{2})\varphi
\end{pmatrix}.
\end{eqnarray}
Here, $\pi=\partial_{t}\varphi$ is the canonical momentum field of
the KG field $\varphi$, which can be discriminated by context from
the number $\pi$. The dagger denotes transposing and complex conjugation
of functional vectors so that $\varphi^{\dagger}j=\int dx\,\bar{\varphi}_{x}j_{x}=\int dk\,\bar{\varphi}_{k}j_{k}/(2\pi)$
in real and Fourier space, respectively. The scalar product of two
component fields $\phi=(\phi^{(\varphi)\dagger},\phi^{(\pi)\dagger})^{\dagger}$
and $\psi=(\psi^{(\varphi)\dagger},\psi^{(\pi)\dagger})^{\dagger}$
is 
\begin{eqnarray}
\phi^{\dagger}\psi & = & \int dx\,\left(\overline{\phi_{x}^{(\varphi)}}\psi_{x}^{(\varphi)}+\overline{\phi_{x}^{(\pi)}}\psi_{x}^{(\pi)}\right),\nonumber \\
 & = & \int\frac{dk}{2\pi}\,\left(\overline{\phi_{k}^{(\varphi)}}\psi_{k}^{(\varphi)}+\overline{\phi_{k}^{(\pi)}}\psi_{k}^{(\pi)}\right)
\end{eqnarray}
in real and Fourier space, respectively.

The KG field dynamics can be derived from the quadratic Hamiltonian
of the dynamical system 
\begin{eqnarray}
\mathcal{H}(\phi) & = & \frac{1}{2}\phi^{\dagger}E\,\phi\label{eq:Hamiltonian}\\
 & = & \int dx\,\frac{1}{2}\left(\pi_{x}^{2}+(\partial_{x}\varphi_{x})^{2}+\mu^{2}\varphi_{x}^{2}\right)\nonumber \\
 & = & \int\frac{dk}{4\pi}\left(|\pi_{k}|^{2}+(\mu^{2}+k^{2})|\varphi_{k}|^{2}\right)\nonumber 
\end{eqnarray}
in abstract, position space and Fourier space notation, respectively.
Here and in the following, $x$ and $y$ are coordinates in position
space, $k$ and $q$ coordinates in continuous or discrete Fourier
space, $t$ is a time coordinate, and coordinate labels determine
in which functional basis a component of a field is to be read out.
The kernel $E$ of the Hamiltonian reads, in the Fourier basis, 
\begin{equation}
E_{kq}=2\pi\delta(k-q)\,\begin{pmatrix}\mu^{2}+k^{2} & 0\\
0 & 1
\end{pmatrix}.
\end{equation}
This determines the KG dynamics via 
\begin{equation}
\partial_{t}\phi=S\:\partial_{\phi}\mathcal{H}(\phi)=S\, E\,\phi,
\end{equation}
with the symplectic matrix 
\begin{equation}
S=\begin{pmatrix}\begin{array}{rr}
0 & 1\\
-1 & 0
\end{array}\end{pmatrix}.\label{eq:Ssymplectiv}
\end{equation}
Therefore, the linear time evolution operator is $L=S\, E$ and the
temporal source is $c=0$ in our example.

The Fourier space representation of the KG dynamics, $(\partial_{t}^{2}+k^{2}+\mu^{2})\varphi_{k}=0$,
has the solution

\begin{eqnarray}
\varphi_{k} & = & a_{k}e^{\iota\omega t}+\overline{a_{-k}}e^{-\iota\omega t}\nonumber \\
\pi_{k} & = & \iota\omega\left(a_{k}e^{\iota\omega t}-\overline{a_{-k}}e^{-\iota\omega t}\right)\label{eq:field-evolution-solution}
\end{eqnarray}
with $\omega=\sqrt{k^{2}+\mu^{2}}$, $\iota=\sqrt{-1}$, and $a_{k}\in\mathbb{C}$.
With respect to the remaining degrees of freedom, the complex amplitudes
$a_{k}$, the Hamiltonian becomes 
\begin{equation}
\mathcal{H}(a)=\int_{0}^{\infty}\frac{dk}{\pi}\:|a_{k}|^{2}\left(k^{2}+\mu^{2}\right)\label{eq:H(a)}
\end{equation}
which implies that these variables are stationary, $\partial_{t}a_{k}=0$.
Therefore, an exact high resolution solution can be specified for
the KG example for all times. This will be compared to approximative
low resolution solutions provided by simulation schemes derived from
IFD and by the usual discretization of differential operators as described
in Appendix \eqref{sub:Discretization-of-differential}. $\blacktriangleleft$

\begin{figure*}
\includegraphics[width=0.5\textwidth]{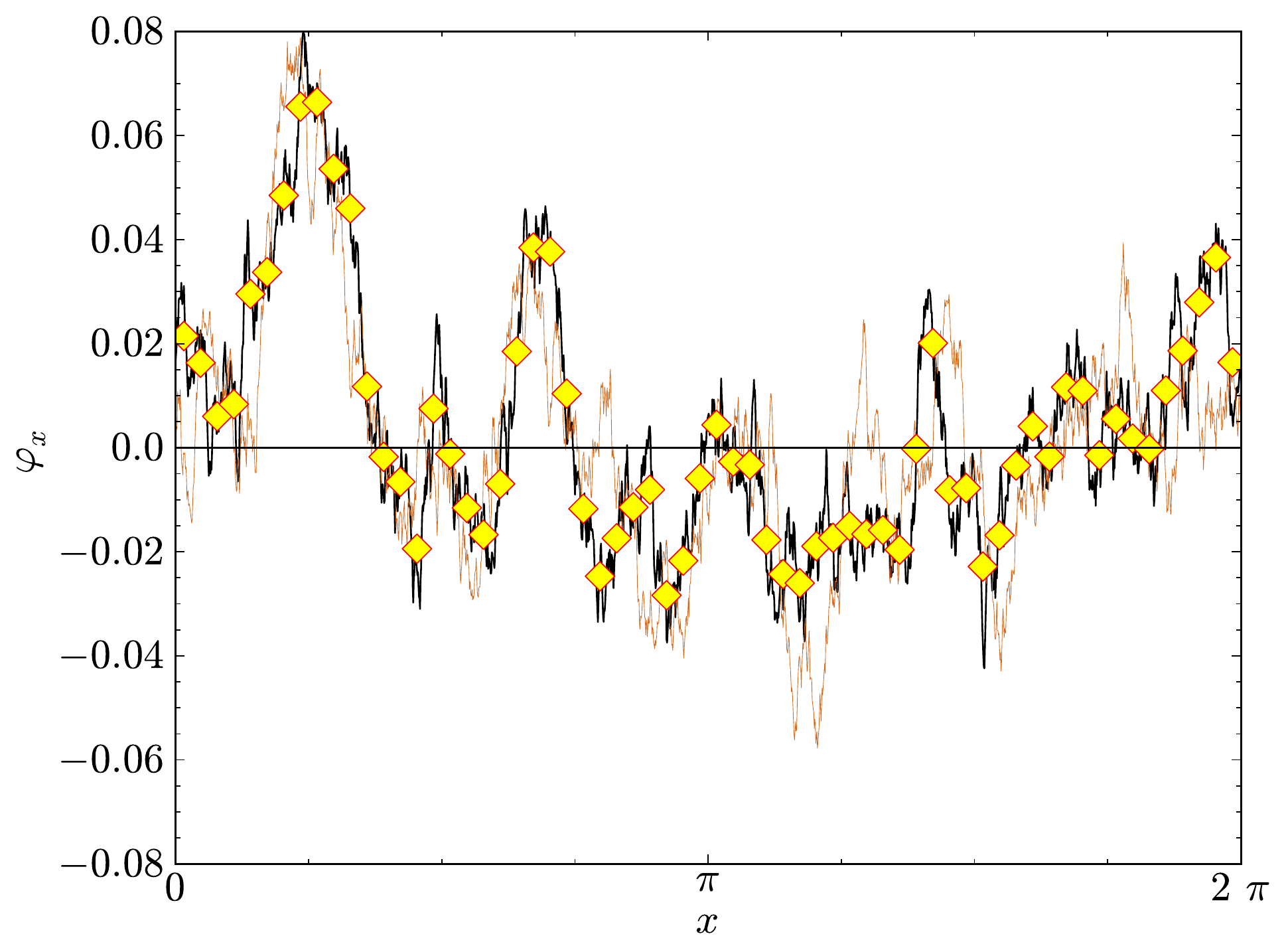}\hspace{-0.5\textwidth}{\small (a)}\hspace{-1.5em}\hspace{0.5\textwidth}\includegraphics[width=0.5\textwidth]{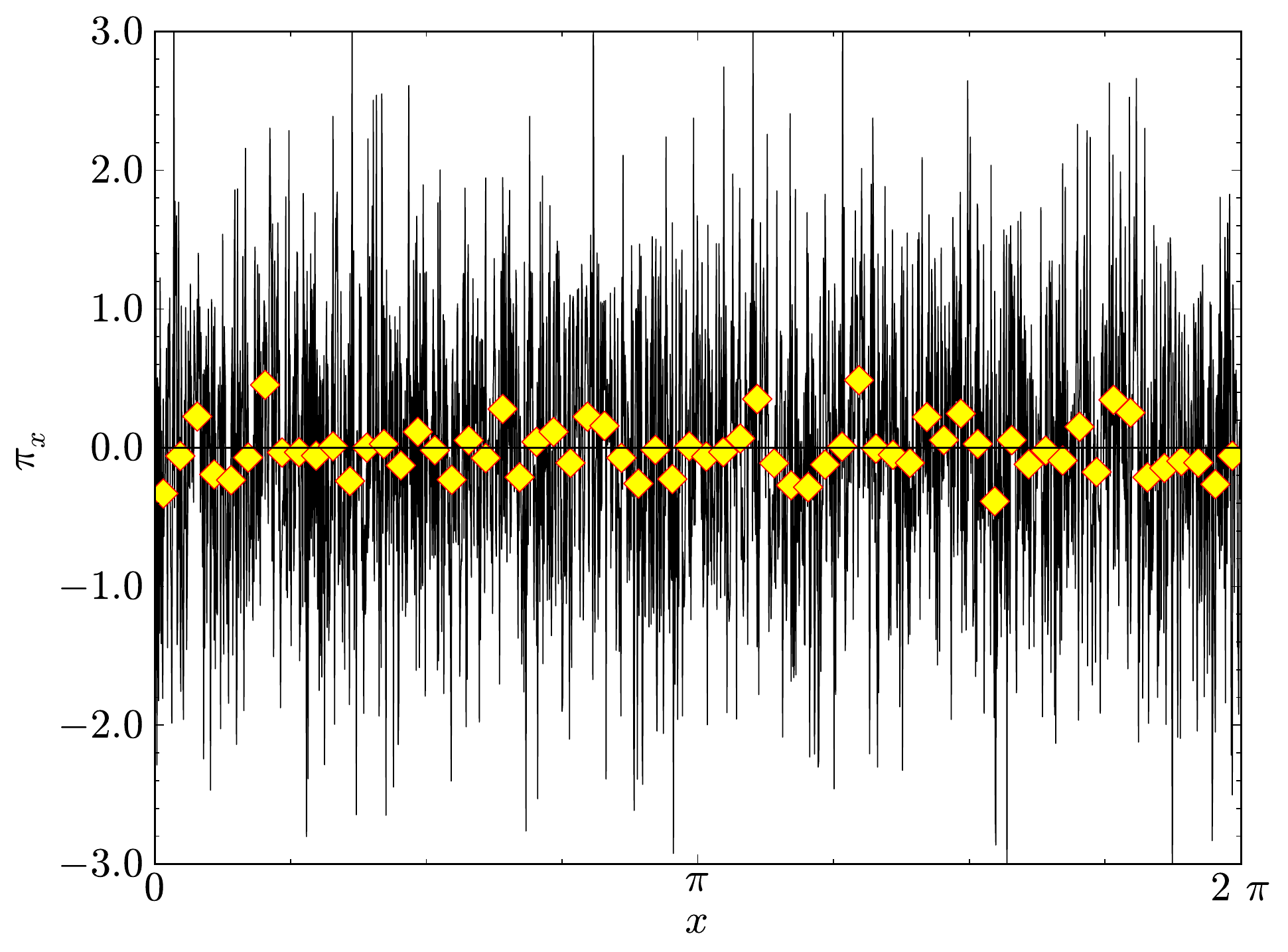}\hspace{-0.5\textwidth}{\small (b)}\hspace{-1.5em}\hspace{0.5\textwidth} 

\caption{(Color online) A realization of a thermally exited KG field $\varphi_{x}$
(a) and its momentum distribution $\pi_{x}$ (b) is shown for $\beta=1$
and $\mu=1$ at $t=0$ with a resolution of 2048 pixels with black
lines passing through the diamond symbols. The low resolution data
with $\mathcal{N}=64$ data points describing the same fields are
shown with yellow diamonds. The field configuration at $t=0.1$ is
also shown in panel (a) with a thin brown (grey) line. The KG field
$\varphi_{x}$ shows a correlated structure due to the suppression
of small scale power by the gradient term in the Hamiltonian, whereas
its momentum field $\pi_{x}$ is just white noise. The loss of small-scale
structure information in the low resolution sampling is especially
apparent for the momentum data. \label{fig:thermal-KG_field}}
\end{figure*}

\subsection{Prior knowledge\label{sub:Prior-knowledge}}

The signal field prior $\p(\phi)$ has to be specified. The prior
should summarize the data-independent knowledge on the field configuration
at current time $t$. For practical reasons, one will typically approximate
it by a Gaussian 
\begin{equation}
\p(\phi)=\mathcal{G}(\phi-\psi,\Phi)\label{eq:Gauss-with-mean}
\end{equation}
with properly chosen mean field $\psi=\left\langle \phi\right\rangle _{(\phi)}$
and prior uncertainty variance $\Phi=\left\langle (\phi-\psi)\text{\,}(\phi-\psi)^{\dagger}\right\rangle _{(\phi)}$.
Such an approximation is often possible, since even non-Gaussian knowledge
states are typically sufficiently well approximated by Gaussians.
Any sophisticated treatment of the otherwise resulting non-linear,
interacting IFT is beyond the scope of this paper.

The Gaussian prior can also be justified from a pure information theoretical
point of view. In case only the prior mean $\psi$ and variance $\Phi$
are known from physical considerations, the MEP distribution of the
field $\phi$ representing exactly this knowledge is given by the
Gaussian \eqref{eq:Gauss-with-mean} with this mean and variance,
as shown in Appendix \ref{sec:Appendix----Maximum}.

Any known mean field $\psi$ can easily be absorbed by the redefinitions
$\phi\rightarrow\phi'=\phi-\psi$ and $c\rightarrow c'=c+L\,\psi$.
This, however, might create a $c$-term even if none existed initially
in the dynamical equation. Therefore we keep the possibility of a
prior mean in the formalism, but note that there is some freedom to
trade a prior mean $\psi$ against a field independent $c$-term and
vice versa.

$\blacktriangleright$ For our illustrative example of a KG field,
we assume that the field was initially in contact and equilibrium
with a thermal reservoir at temperature $\beta^{-1}$ and became decoupled
from it at some time $t_{0}=0$. The initial probability function
of the field is therefore thermal, 
\begin{equation}
\p(\phi|\beta)=\frac{1}{Z_{\beta}}e^{-\beta\,\mathcal{H}(\phi)}=\prod_{k}\frac{1}{z_{k}}e^{-2\beta\,|a_{k}|^{2}\left(k^{2}+\mu^{2}\right)}.\label{eq:thermal prior}
\end{equation}
It separates into independently excited modes, which do not exchange
energy at later times because the amplitudes are stationary. Thus,
an initially established thermal state stays thermal and at the same
temperature for all times. The partition function is given by a complex
Gaussian integral for each mode and is 
\begin{equation}
Z_{\beta}\equiv\int\D\phi\, e^{-\beta\,\mathcal{H}(\phi)}=\prod_{k}\underbrace{\frac{\pi}{2\beta\,(k^{2}+\mu^{2})}}_{z_{k}},
\end{equation}
where the product goes over all accessible positive wave vectors.

Since the energy Hamiltonian $\mathcal{H}(\phi)=\frac{1}{2}\phi^{\dagger}E\,\phi$
is quadratic in $\phi$, the prior information Hamiltonian $H(\phi|\beta)=\beta\,\mathcal{H}(\phi)=\frac{\beta}{2}\phi^{\dagger}E\,\phi$
is quadratic as well. The prior is simply a Gaussian $\p(\phi|\beta)=\mathcal{G}(\phi,\Phi)$
with zero mean $\psi=0$ and covariance $\Phi=(\beta\, E)^{-1}$.
In Fourier space this reads 
\begin{equation}
\Phi_{kq}=\frac{2\pi}{\beta}\delta(k-q)\,\begin{pmatrix}\left(\mu^{2}+k^{2}\right)^{-1} & 0\\
0 & 1
\end{pmatrix}\label{eq:Phikq}
\end{equation}
and in position space it is 
\begin{equation}
\Phi_{xy}=\frac{1}{\beta}\:\begin{pmatrix}\frac{1}{2\mu}\, e^{-\mu\,|x-y|} & 0\\
0 & \delta(x-y)
\end{pmatrix}.\label{eq:correlation_phi}
\end{equation}
A KG field realization drawn from \eqref{eq:thermal prior} for $\beta=1$
and $\mu=1$ is displayed in Fig. \eqref{fig:thermal-KG_field}. There,
the different spatial correlation structures of the field values with
$\left\langle \varphi_{x}\varphi_{y}\right\rangle _{(\phi)}=(2\mu\beta)^{-1}e^{-\mu|x-y|}$
and field momenta with $\left\langle \pi_{x}\pi_{y}\right\rangle _{(\phi)}=\beta{}^{-1}\delta(x-y)$,
as given by \eqref{eq:correlation_phi}, can be seen. $\blacktriangleleft$

\subsection{Data constraints}

In addition to the relatively vague prior knowledge, the field is
constrained by the finite dimensional data vector $d=(d_{i})_{i}$
in computer memory. The data is assumed to represent linear statements
on the field of the form $d=Rs+n,$ c.f. Eq. \eqref{eq:D=00003D00003D00003DRs+n}.
In typical numerical simulation schemes, the response operator might
just express an averaging of the field within some environment $\Omega_{i}$
of a grid point $x_{i}\in\Omega_{i}$, i.e. 
\begin{equation}
R_{ix}=\frac{1}{|\Omega_{i}|}\theta(x\in\Omega_{i}),\label{eq:Rix}
\end{equation}
where the logical theta function 
\begin{equation}
\theta(x\in\Omega_{i})=\p(x\in\Omega_{i}|x,\Omega_{i})=\begin{cases}
1\, & x\in\Omega_{i}\\
0\, & x\notin\Omega_{i}
\end{cases}
\end{equation}
is one, if the condition in its argument is true, otherwise it is
zero. In schemes based on grid cells or space tessellations, the grid
point volumes are disjoint, $\Omega_{i}\cap\Omega_{j}=\emptyset$
for $i\neq j$. In case a conserved quantity should be conserved as
accurately as possible, the total amount of the quantity within the
cells of a space tessellation as well as the currents of the quantity
through the surfaces of the tessellation cells might be used as data.
In smoothed particle hydrodynamics, the volumes overlap and are usually
also structured by radially declining kernel functions that have evolving
locations and sizes.

For the moment, we only have to deal with the data at one instant,
and need only to know that it depends linearly on the underlying field
by a known relation of the form $d=R\phi+n$. This relation might
or might not be the same at the next instant, depending on the design
choices for $R=R_{t}$ (stationary grid or Lagrangian moving mesh).
$R_{t}$ could even be determined by the IFD formalism itself by requiring
minimal information loss of the scheme, as we will do later for the
KG field example in Sect. \eqref{sub:Data-update}.

The simulation data vector $d$ can even be extended also to contain
measurement data on the system to be simulated (e.g. the weather)
obtained for the current simulation time. If this auxiliary data $\mathfrak{d}$
resulted from a linear measurement $\mathbb{\mathfrak{d=R}}\phi+\mathfrak{n}$
with response $\mathfrak{R}$ and Gaussian noise $\mathfrak{n}$ with
covariance $\mathfrak{N}$, only the replacements 
\begin{equation}
d\rightarrow\left(\begin{array}{c}
d\\
\mathfrak{d}
\end{array}\right),\; R\rightarrow\left(\begin{array}{c}
R\\
\mathfrak{R}
\end{array}\right),\;\mathrm{\mbox{and }}N\rightarrow\left(\begin{array}{cc}
N & 0\\
0 & \mathfrak{N}
\end{array}\right)\label{eq:dataAssimilation}
\end{equation}
are needed.%
\footnote{The block diagonal structure of the extended noise covariance matrix
assumes that the measurement error and the simulation error are uncorrelated.
This assumption would be improper in case repeated measurements with
the same incorrectly calibrated instrument are assimilated into the
simulation. In that case, correlations among the simulation and measurement
data errors could exist since the correlated measurement errors are
partly imprinted onto the simulation data. %
} This way, the measurement information is assimilated into the simulation
scheme and can be evolved into the future (or into the past, if the
simulation is backward in time). The added data could become simulation
degrees of freedom, or they could be discarded at the next simulation
time step after their information was transferred to the simulation
data via the entropic matching operation. The former option would
certainly conserve more information, the latter is somehow similar
to what is done in particle filter methods as described in Appendix
\ref{sub:Data-assimilation-methods}.

The ensemble of field configurations constrained by the data via \eqref{eq:D=00003D00003D00003DRs+n}
and by the prior via \eqref{eq:Gauss-with-mean} is then 
\begin{equation}
\p(\phi|d)=\mathcal{G}(\phi-m,\, D),\label{eq:posterior}
\end{equation}
where 
\begin{eqnarray}
D & = & \left(\Phi^{-1}+R^{\dagger}N^{-1}R\right)^{-1}\quad\mbox{and}\nonumber \\
m & = & \psi+W\,\left(d-R\,\psi\right)=D\,\left(R^{\dagger}N^{-1}d+\Phi^{-1}\psi\right).\label{eq:mD posterior}
\end{eqnarray}
The mean is shifted here with respect to \eqref{eq:m=00003D00003D00003DWd}
due to the non-vanishing prior mean $\psi$.

In case that external data $\mathfrak{d}$ is to be assimilated into
the simulation, applying replacements of \eqref{eq:dataAssimilation}
to \eqref{eq:mD posterior} and expanding this yields $D=\left(\Phi^{-1}+R^{\dagger}N^{-1}R+\mathfrak{R}^{\dagger}\mathfrak{N}^{-1}\mathfrak{R}\right)^{-1}$
and $d=D\,\left(R^{\dagger}N^{-1}d+\mathfrak{R}^{\dagger}\mathfrak{N}^{-1}\mathfrak{d}+\Phi^{-1}\psi\right)$.
Thus, data assimilation is very naturally done in IFD since simulation
and measurement data shape the field posterior $\p(\phi|d)=\mathcal{G}(\phi-m,\, D)$
in a similar way.

$\blacktriangleright$ In our example of the KG field we want to deal
with the simplest possible data as given by \eqref{eq:D=00003D00003D00003DRs+n}
and \eqref{eq:Rix} that lives on a regular grid, with equidistant,
space filling and disjoint pixel volumes $\Omega_{i}=[i\,\Delta,\,(i+1)\,\Delta)$,
with $\Delta>0$ being the grid spacings. Since on a computer one
can only deal with finite domains, we assume periodic boundary conditions
for the interval $\Omega=\cup_{i}\Omega_{i}=[0,\,2\pi]$ and require
that the number of grid points $\mathcal{N}=2\pi/\Delta\in\mathbb{N}$.
The Fourier transformed field is then 
\begin{eqnarray}
\phi_{k} & = & \int_{0}^{2\pi}dx\, e^{\iota kx}\phi_{x},\quad\mbox{with}\\
\phi_{x} & = & \sum_{k=-\infty}^{\infty}\frac{1}{2\pi}\, e^{-\iota kx}\phi_{k}.
\end{eqnarray}
Here the following substitution with respect to the infinitely extended
case have been made: $\int\, dx\rightarrow\int_{0}^{2\pi}dx$ and
$\int\frac{dk}{2\pi}\rightarrow\sum_{k=-\infty}^{\infty}\frac{1}{2\pi}$,
which are the appropriately weighted sums of the scalar products in
position and Fourier space, respectively. Furthermore, we note that
$\delta(k-q)\rightarrow\delta_{kq}$ in this case, so that the unit
operator is $\mathbb{1}_{kq}=2\pi\delta_{kq}$ and the field covariance
\eqref{eq:Phikq} reads 
\begin{equation}
\Phi_{kq}=\frac{2\pi}{\beta}\delta_{kq}\,\begin{pmatrix}\left(\mu^{2}+k^{2}\right)^{-1} & 0\\
0 & 1
\end{pmatrix}.\label{eq:Phikq-1}
\end{equation}

Since the data space is finite, its Fourier space is also finite,
where 
\begin{eqnarray}
d_{k} & = & \sum_{i=0}^{\mathcal{N}-1}\Delta\, e^{\iota ki\Delta}d_{i},\quad\mbox{with}\\
d_{i} & = & \sum_{k=0}^{\mathcal{N}-1}\frac{1}{2\pi}\, e^{-\iota ki\Delta}d_{k},
\end{eqnarray}
and $k\in\{0,\,\ldots\:\mathcal{N}-1\}$. Higher or negative Fourier
modes do not carry any additional information due to the Nyquist theorem.%
\footnote{These conventions for the discrete Fourier transformation might appear
a bit unusual, but they have the advantage that they match best the
continuous space Fourier convention used in physics. They permit us
to use all derived Fourier space equations for the KG field without
changing normalization constants and with the intuitive identifications
$dx\rightarrow\Delta$, $x\rightarrow i\Delta$ and $k\rightarrow k$.%
}

The Fourier transformed response, 
\begin{eqnarray}
R_{kq} & = & 2\,\pi\,\theta(q-k\in\mathcal{N}\,\mathbb{Z})\,\frac{1-e^{-\iota q\Delta}}{\iota q\Delta}\label{eq:Rkq}\\
 & = & 2\,\pi\,\theta(q-k\in\mathcal{N}\,\mathbb{Z})\, e^{-\frac{1}{2}\iota q\Delta}\mathrm{sinc}\left(\frac{1}{2}q\Delta\right),
\end{eqnarray}
is block diagonal in the reduced Fourier space of the data with $k\in\{0,\,\ldots\ \mathcal{N}-1\}$.
Note, however, that higher Fourier modes of the field $\phi_{q}$
with $q\in k+\mathcal{N}\,\mathbb{Z}$, which carry information on
sub-grid structure, imprint also onto the data and blend with the
lower Fourier modes $k\in\{0,\,\ldots\ \mathcal{N}-1\}$. Therefore
a unique reconstruction of the individual Fourier modes from the data
alone is impossible even within the range $q\in\{0,\,\ldots\ \mathcal{N}-1\}$.

The individual terms in \eqref{eq:Rkq} can easily be understood.
The $\exp(-\frac{1}{2}\iota q\Delta)$ term stems from the fact that
the centers of the pixel volumes are shifted by $\frac{1}{2}\Delta$
from the pixel positions $i\Delta$ used in the definition of the
Fourier transformation. The sinc-function is the Fourier space transform
of the pixel window. It encodes how well a given Fourier mode is represented
in the data, and therefore how well it is protected from noise and
confusion with other modes imprinted onto the same data mode.

The data space signal covariance, which is needed by the Wiener filter,
is%
\footnote{Here, we used the following identities: 
\[
\sum_{i\in\mathbb{Z}}\frac{1}{(a+i)^{2}}=\frac{\pi^{2}}{\sin^{2}(\pi a)}
\]
and 
\begin{eqnarray*}
 &  & \sum_{i\in\mathbb{Z}}\frac{1}{(a+i)^{2}((a+i)^{2}+b^{2})}=\\
 &  & \frac{\pi}{b^{3}}\left[\frac{b\pi}{\sin^{2}(\pi a)}-\frac{\sinh(2\pi b)}{\mathrm{cosh}(2\pi b)-\cos(2\pi a)}\right].
\end{eqnarray*}
} 
\begin{eqnarray}
\tilde{\Phi}_{kq} & = & \left(R\,\Phi\, R^{\dagger}\right)_{kq}=\begin{pmatrix}\tilde{\Phi}_{kq}^{(\varphi)} & 0\\
0 & \tilde{\Phi}_{kq}^{(\pi)}
\end{pmatrix},\;\mbox{with}\\
\tilde{\Phi}_{kq}^{(\varphi)} & = & \frac{\tilde{\Phi}_{kq}^{(\pi)}}{\mu^{2}}\,\begin{cases}
1 & k=0\\
\left[1-\frac{2}{\mu\Delta}\:\frac{\sinh(\mu\Delta)\sin^{2}(\frac{1}{2}k\Delta)}{\cosh(\mu\Delta)-\cos(k\Delta)}\right] & k\neq0
\end{cases},\nonumber \\
\tilde{\Phi}_{kq}^{(\pi)} & = & \frac{2\pi\delta_{kq}}{\beta}\begin{cases}
1 & k=0\\
\frac{1-\cos(k\Delta)}{2\sin^{2}(\frac{1}{2}k\Delta)} & k\neq0
\end{cases}\;.\nonumber 
\end{eqnarray}

Since the field covariance and response are translationally invariant
we have every reason to believe that the noise statistics, which are
fed only by approximation errors depending on these latter two quantities,
will also be translationally invariant in data space. Therefore its
covariance will also be diagonal in discrete Fourier space: 
\begin{equation}
N_{kq}=2\pi\delta_{kq}\begin{pmatrix}\eta_{k}^{(\varphi)} & \eta_{k}^{(\mathrm{c})}\\
\overline{\eta_{k}^{(\mathrm{c})}} & \eta_{k}^{(\pi)}
\end{pmatrix},
\end{equation}
where $\eta^{(\varphi)}$, $\eta^{(\pi)}$, and $\eta^{(\mathrm{c})}$
are the noise spectra of the field value data, the field momenta data,
and the cross-spectra of those, respectively. However, in Sect. \eqref{sub:Data-update}
we will show that the ideal IFD scheme stays noiseless if it was initially
noiseless. Therefore we can set $N\rightarrow0$ for all times and
use the $\eta$-parameters to ensure consistency of all formula. They
will be set to zero at the end of the calculation if this is a permitted
limit.

Taking the noiseless case as granted for the moment, the Wiener filter
becomes 
\begin{eqnarray}
W_{kq} & = & \left(\Phi\, R^{\dagger}\tilde{\Phi}^{-1}\right)_{kq}\\
 & = & 2\pi\theta(q=k\,\mathrm{mod}\,\mathcal{N})\, e^{\frac{1}{2}\iota k\Delta}\mathrm{sinc}\left(\frac{1}{2}k\Delta\right)\,\times\nonumber \\
 &  & \frac{2\sin^{2}\left(\frac{1}{2}q\Delta\right)}{1-\cos(q\Delta)}\,\times\nonumber \\
 &  & \begin{pmatrix}\frac{\mu^{2}}{\mu^{2}+k^{2}}\left[1-\frac{2}{\mu\Delta}\:\frac{\sinh(\mu\Delta)\,\sin^{2}(\frac{1}{2}k\Delta)}{\cosh(\mu\Delta)-\cos(k\Delta)}\right]^{-1} & 0\\
0 & 1
\end{pmatrix}.\nonumber 
\end{eqnarray}
For a reconstructed signal image generated by this Wiener filter,
any image Fourier mode $k\in\mathbb{Z}$ gets exited by its first
Brillouin zone data space mode $q=k\,\mathrm{mod}\,\mathcal{N}\in\{0,\ldots\mathcal{N}-1\}$.
Thereby, all Fourier modes $k\in\mathbb{Z}$ of the mean field $m=W\, d$
get some non-trivial value if the corresponding data mode $k\,\mathrm{mod}\,\mathcal{N}$
was non-zero. $\blacktriangleleft$

\subsection{Field evolution\label{sub:Field-evolution}}

A Gaussian knowledge state $\p(\phi|t)=\p(\phi|\, d=d(t))=\mathcal{G}(\phi-m,\, D)$
at some initial time $t$ is represented by the data $d=d_{t}$, which
determinesthe mean field via $m=W\, d$. The field uncertainty variance
$D$ is data- and time-independent in our example, but not in general.
The knowledge state $\mathcal{P}(\phi|t)$ has to be evolved to a
infinitesimally later time $t'=t+\delta t$ via the evolution of the
individual field configurations.

An individual field configuration $\phi=\phi_{t}$ at initial time
$t$ becomes $\phi'=\phi_{t'}\,\widehat{=}\,\phi_{t}+\delta t\:\dot{\phi}_{t}=\phi_{t}+\delta t\,(L\,\phi_{t}+c)$,
where the time derivative is given by \eqref{eq:linear dynamics}.
Here, and in the following, we drop non-essential terms of $\mathcal{O}(\delta t^{2})$,
as indicated by ``$\widehat{=}$''. The time-evolved knowledge state
therefore becomes 
\begin{equation}
\p(\phi'|\, d)=\p(\phi|\, d)\left|\frac{\partial\phi}{\partial\phi'}\right|
\end{equation}
by conservation of probability density. We need to calculate the Jacobian
up to linear order in $\delta t$. This is most simply done from the
inverse Jacobian, 
\begin{eqnarray}
\left|\frac{\partial\phi'}{\partial\phi}\right| & = & \left|\mathbb{1}+\delta t\, L\right|=\exp\log\left|\mathbb{1}+\delta t\, L\right|\nonumber \\
 & \widehat{=} & \exp\mathrm{Tr}\left(\delta t\, L\right)\,\widehat{=}\,\mathrm{1+\delta t\, Tr}\left(L\right).
\end{eqnarray}

In case of a linear Hamiltonian dynamics $\partial_{t}\phi=S\,\partial_{\phi}\mathcal{H}(\phi)$,
with dynamical Hamiltonian of the from $\mathcal{H}(\phi)=\frac{1}{2}\phi^{\dagger}E\,\phi+b^{\dagger}\phi$
and $E$ being block diagonal in the field value $\varphi$ and field
momentum $\pi$ eigenspaces, we have $L=S\, E$ and $c=S\, b$. The
Jacobian is then unity, since 
\begin{eqnarray}
\mathrm{Tr}\left(L\right) & = & \mathrm{Tr}\left(S\, E\right)=\mathrm{Tr}\left(\begin{pmatrix}0 & 1\\
-1 & 0
\end{pmatrix}\,\begin{pmatrix}E^{(\phi)} & 0\\
0 & E^{(\pi)}
\end{pmatrix}\right)\nonumber \\
 & = & \mathrm{Tr}\begin{pmatrix}0 & -E^{(\pi)}\\
E^{(\phi)} & 0
\end{pmatrix}=0.
\end{eqnarray}
This is not surprising, since it is well known that symplectic Hamiltonian
systems conserve the phase space density, so that the unity of the
Jacobian is also valid for non-infinitesimal time steps $\delta t$
in such cases.

In general, for non-Hamiltonian systems, the Jacobian can be different
from one. It can be larger for systems with dynamical attractors or
with dissipation (Navier-Stokes equations) and it can be smaller for
systems with diverging phase-space flows, like chaotic inflation in
cosmology or driven hydrodynamical turbulence (without significant
dissipation).

The evolved knowledge state, or the knowledge state on the evolved
field, is therefore 
\begin{eqnarray}
\p(\phi'|\, d) & \widehat{=} & \p(\phi=\text{\ensuremath{\phi}}'-\delta t\,\dot{\phi}|\, d)\,|\partial\phi/\partial\phi'|\label{eq:evolved state}\\
 & \widehat{=} & \mathcal{G}(\phi'-\delta t\,(L\phi'+c)-m,\, D)\,(\mathrm{1-\delta t\, Tr}\left(L\right))\nonumber \\
 & \widehat{=} & \mathcal{G}(\phi'-m^{*},\, D^{*}),\nonumber 
\end{eqnarray}
with%
\footnote{The key to understand this result is a short rearrangement in the
exponent of the Gaussian, 
\begin{eqnarray*}
 &  & ((1-\delta t\, L)\phi'-m-\delta t\, c)^{\dagger}D^{-1}((1-\delta t\, L)\phi'-m-\delta t\, c)\\
 &  & =(\phi'-\underbrace{\left((1-\delta t\, L)^{-1}m+\delta t\, c\right)}_{m^{*}})^{\dagger}\underbrace{(1-\delta t\, L)^{\dagger}D^{-1}(1-\delta t\, L)}_{D^{*-1}}\,(\phi'-m')
\end{eqnarray*}
the $\delta t$-expansion of the new mean field 
\[
m^{*}\widehat{=}(1+\delta t\, L)\, m+\delta t\, c,
\]
that of the new uncertainty variance 
\begin{eqnarray*}
D^{*} & \widehat{=} & (1+\delta t\, L)\, D\,(1+\delta t\, L)^{\dagger}\\
 & \widehat{=} & D+\delta t\,(L\, D+D\, L^{\dagger}),
\end{eqnarray*}
and its determinant 
\[
|D^{*}|\;\widehat{=}\;(\mathrm{1+2\delta t\, Tr}\left(L\right))\,|D|.
\]
} 
\begin{eqnarray}
m^{*} & \widehat{=} & m+\delta t(c+L\, m)\label{eq:m'D'}\\
 & \widehat{=} & (1+\delta t\, L)\,\left(\psi+W\,\left(d-R\,\psi\right)\right)+\delta t\, c,\nonumber \\
D^{*} & \widehat{=} & D+\delta t\,(L\, D+D\, L^{\dagger}).\nonumber \\
D^{*-1} & \widehat{=} & D^{-1}-\delta t\,(D^{-1}L+L^{\dagger}D^{-1}).\nonumber 
\end{eqnarray}

$\blacktriangleright$ In case of our KG field, we have $\mathrm{Tr}\left(L\right)=0$
due to the symplectic dynamics with $L=S\, E$ and $c=0$, as well
as $m^{*}\:\widehat{=}\: m+\delta t\, S\, E\, m$. Furthermore, using
$L=S\, E$, $S^{\dagger}=-S$ , $D^{-1}=\Phi^{-1}+R^{\dagger}N^{-1}R$,
and $\Phi^{-1}=\beta\, E$, we get $D^{*-1}\,\widehat{=}\, D^{-1}-\delta t\,(R^{\dagger}N^{-1}R\, S\, E-E\, S\, R^{\dagger}N^{-1}R)$.

The evolved mean field still can be regarded to be parametrized by
the data, however, in a different way, $m^{*}=(1+\delta t\, S\, E)\, W\, d$.
It is not clear in general whether a new dataset $d'$ can be found
that expresses this new mean field via the original parametrization
$m'=W\, d'$ (or with the appropriate $W'$, in case that also $D'$
changed). This is because the functional forms of the two parametrizations
differ since $W$ and $L=S\, E$ operate on completely different vector
spaces, the discrete data space and the continuous field space, respectively.

Therefore entropic matching will be used to choose a $d'$ that determines
$\p'(\phi'|\, d')$ such that it captures most of the information
content of $\p(\phi'|\, d)$. $\blacktriangleleft$

\subsection{Prior update\label{sub:Prior-update}}

The field prior for time $t'$ has to be updated since the sub-grid
statistics might have changed. For example some of the energy contained
in sub-grid modes might dissipate, leading to a different $\p(\phi')=\mathcal{G}(\phi'-\psi',\Phi')$
as parametrized via the updated prior mean $\psi'$ and variance $\Phi'$.

$\blacktriangleright$ In case of our KG field, energy conservation
of the dynamics leads to an unchanged prior for the evolved field
$\p(\phi')=\mathcal{G}(\phi',\Phi)$, still with $\Phi=(\beta\, E)^{-1}$.
$\blacktriangleleft$

\subsection{Data update\label{sub:Data-update}}

The new data has to be determined from its relation to the updated
field. Again, we assume the new data to depend linearly on the evolved
field 
\[
d'=R'\,\phi'+n'.
\]

Note that we could chose a different pixilation at $t'$, leading
to a different response $R'$, propagator $D'$, and Wiener filter
$W'.$ This is needed e.g. in case a simulation with moving or adaptive
mesh is to be developed. It can even be considered that the response
operator determination becomes a part of the entropic matching step,
leading to an information optimal moving mesh.

Furthermore, we have to allow for a changed noise level, with new
covariance $N'$, since the meaning of the data values could have
changed with changed pixilation and since we might have to allow for
additional uncertainty in order to capture any mismatch between the
new parametrized posterior and the evolved field posterior.

According to \eqref{eq:posterior} and \eqref{eq:mD posterior} the
relation of new posterior and new data is 
\begin{equation}
\p(\phi'|d')=\mathcal{G}(\phi'-m',\, D'),\label{eq:posterior-1}
\end{equation}
where 
\begin{eqnarray}
D' & = & \left(\Phi'^{-1}+R'^{\dagger}N'^{-1}R'\right)^{-1},\nonumber \\
m' & = & \psi'+W'\,\left(d'-R'\,\psi'\right)\nonumber \\
 & = & D'\,\left(R'^{\dagger}N'^{-1}d'+\Phi'^{-1}\psi'\right),\quad\mbox{and}\nonumber \\
W' & = & D'\, R'^{\dagger}N'^{-1}=\Phi'R'^{\dagger}(\underbrace{R'\,\Phi'\, R'^{\dagger}}_{\tilde{\Phi}'}+N')^{-1}.
\end{eqnarray}

Now, the new posterior $\p'=\p(\phi'|d')$ should match the evolved
posterior $\p=\p(\phi'|d)$ as well as possible. According to \eqref{eq:entropy}
the cross entropy of the former with the latter is 
\begin{eqnarray}
\mathcal{S}(\p'|\p) & = & -\frac{1}{2}\mathrm{Tr}\left[\left(\delta m\,\delta m^{\dagger}+D'\right)\, D^{*-1}+\right.\nonumber \\
 &  & \left.\mathbb{1}+\log(D'\, D^{*-1})\right]
\end{eqnarray}
with $\delta m=m'-m^{*}$.

Maximizing this entropy with respect to the new data $d'$ yields
\begin{eqnarray}
-\partial_{d'}\mathcal{S} & = & (\partial_{d'}m')^{\dagger}D^{*-1}\delta m\nonumber \\
 & = & W'^{\dagger}D^{*-1}(W'\,\left(d'-R'\,\psi'\right)+\psi'-m^{*})\:=\:0\nonumber \\
\Rightarrow d' & = & R'\,\psi'+\label{eq:d' eq}\\
 &  & \left(W'^{\dagger}D^{*-1}W'\right)^{-1}W'^{\dagger}D^{*-1}\left(m^{*}-\psi'\right).\nonumber 
\end{eqnarray}

This is the general formula to update the data. It should be expanded
up to linear order in all the relevant changes in response $R'=R+\delta R$,
noise covariance $N'=N+\delta N$, and prior parameters $\Phi'=\Phi+\delta\Phi$
and $\psi'=\psi+\delta\psi$, as well as in time $t'=t+\delta t$.
The resulting general formula is lengthy and not directly instructive%
\footnote{A few useful identities, when dealing with \eqref{eq:d' eq} might
be in order. A short calculation shows that up to linear order in
$\delta t$ 
\begin{eqnarray*}
 &  & \left(W'^{\dagger}D^{*-1}W'\right)^{-1}W'^{\dagger}\\
 & = & (\tilde{\Phi}'+N')\,(R'\Phi'D^{*-1}\Phi'R^{\dagger})^{-1}\, R'\Phi'\\
 & \widehat{=} & (\tilde{\Phi}'+N')\,(R'\Phi'\left(D^{-1}-\delta t\,(D^{-1}L+L^{\dagger}D^{-1})\right)\Phi'R'^{\dagger})^{-1}\, R'\Phi'\\
 & \widehat{=} & (\tilde{\Phi}'+N')\,(\tilde{D}+\delta t\,\tilde{D}\, R'\Phi'(D^{-1}L+L^{\dagger}D^{-1})\Phi'R'^{\dagger}\tilde{\, D\,})\, R'\Phi',
\end{eqnarray*}
with $\tilde{D}=(R'\Phi'D^{-1}\Phi'R^{\dagger})^{-1}$and that 
\begin{eqnarray*}
 &  & D^{*-1}\left(m^{*}-\psi'\right)\\
 & \widehat{=} & \left(D^{-1}-\delta t\,(D^{-1}L+L^{\dagger}D^{-1})\right)\,\left((1+\delta t\, L)\,\left(\psi+W\,\left(d-R\,\psi\right)\right)+\delta t\, c\right).\\
 & \widehat{=} & D^{-1}\psi+R^{\dagger}N^{-1}(d-R\psi)\\
 &  & +\delta t\left[D^{-1}c-L^{\dagger}\left(D^{-1}\psi+R^{\dagger}N^{-1}\left(d-R\,\psi\right)\right)\right].
\end{eqnarray*}
}, therefore we concentrate here more on special cases.

The update of the uncertainty variance is also obtained by maximizing
the entropy with respect to the degrees of freedom of $D'=(\Phi'+R'^{\dagger}N'^{-1}R')^{-1}$.
These could be the location of the new pixel positions, which influence
$R'$, or an updated noise level, influencing $N'$, or properties
of the field prior expressed via $\Phi'$ and $\psi'$.

We combine these degrees of freedom into the single vector $\eta$,
irrespective of whether they determine $R'$, $N'$, $\Phi',$ $\psi'$,
or combinations thereof. The entropic matching of the updated uncertainty
variance $D'=D(\eta+\delta\eta)=D(\eta)+\sum_{i}\delta\eta_{i}\Gamma_{i}+\mathcal{O}(\delta\eta^{2})$,
with $\Gamma_{i}=\partial_{\eta_{i}}D(\eta)$ the linear changes due
to changes in the degrees of freedom, is then given by 
\begin{eqnarray}
-\partial_{\eta}\mathcal{S} & = & \frac{1}{2}\mathrm{Tr}\left[\left(\partial_{\eta}D'\right)\,\left(D^{*-1}-D'^{-1}\right)\right]=0\nonumber \\
\Rightarrow\delta\eta & = & C^{-1}b,\ \mbox{with}\label{eq:D' eq}\\
b_{i} & = & \mathrm{Tr}\left[\Gamma_{i}\,\left(D^{*-1}-D{}^{-1}\right)\right]\ \mbox{and}\nonumber \\
C_{ij} & = & \mathrm{Tr}\left[\Gamma_{i}\, D{}^{-1}\,\Gamma_{j}\, D{}^{-1}\right].\nonumber 
\end{eqnarray}
From the first line it is already apparent, that if $D'$ is able
to match $D^{*}$ exactly, then it will do so. The detailed formula
for updating response, noise, and prior can be complex, since operator
inversions are involved. In general, approximations might be necessary
here in order to proceed with a reasonable computational complexity.

The formula \eqref{eq:d' eq} and \eqref{eq:D' eq} form the desired
simulation scheme. The scheme deals optimally with time dependent
pixilation, non-Hamiltonian dynamics, sub-grid processes, as well
as with the accumulation of discretization errors. The price of this
generality is a higher complexity of the detailed formula compared
to many ad-hoc schemes. These formula have to be analyzed case by
case to identify the optimal numerical implementation strategy. In
order to show this in a simple example, we turn again to the KG field.

$\blacktriangleright$ Assuming that we have all freedom to chose
$R',$ $N',$ and $\Phi'$ to match $D'^{-1}=\Phi'^{-1}+R'^{\dagger}N'^{-1}R'$
exactly with 
\begin{eqnarray}
D^{*-1} & \widehat{=} & \Phi^{-1}+\\
 &  & R^{\dagger}N^{-1}R-\delta t\,(R^{\dagger}N^{-1}R\, S\, E-E\, S\, R^{\dagger}N^{-1}R)\nonumber 
\end{eqnarray}
as derived in Sect. \ref{sub:Field-evolution}, we would immediately
use $\Phi'=\Phi$ and try to accommodate the change in variance in
a changed response or noise. Thus the unchanged signal covariance
also results from the data update via the MEP. The considerations
to update the prior in Sect. \ref{sub:Prior-update} were therefore
superfluous in this case. The updated prior mean $\psi'$ could also
be derived by maximizing the entropy with respect to it. It is not
surprising that it turns out to be $\psi'=\psi=0$.

Writing $R'=R+\delta R$ and $N'=N+\delta N$ we find 
\begin{eqnarray}
D'^{-1} & \widehat{=} & \Phi+R^{\dagger}N^{-1}R+\delta R^{\dagger}N^{-1}R+\nonumber \\
 &  & R^{\dagger}N^{-1}\delta R-R^{\dagger}N^{-1}\delta N\, N^{-1}R.
\end{eqnarray}
Comparing the terms of the last two equations, we conclude that the
best match is found by the identification 
\begin{eqnarray}
\delta R & = & -\delta t\, R\, S\, E,\nonumber \\
\delta N & = & 0.
\end{eqnarray}
Thus, the noise should stay unchanged and can be assumed to be zero
for all times it was zero initially, which we will assume in the following.
The response of an optimal scheme should however evolve according
to $\partial_{t}R_{t}=-R_{t}\, S\, E$. This can actually be solved
analytically, providing 
\begin{equation}
R_{t}=R\, T_{-t},
\end{equation}
with the time translation operator 
\begin{eqnarray}
 &  & (T_{t})_{k\, q}=\left(e^{S\, E\, t}\right)_{k\, q}=\label{eq:T_time-translation}\\
 &  & \mathbb{1}_{k\, q}\left[\cos(\omega_{k}t)\,\begin{pmatrix}1 & 0\\
0 & 1
\end{pmatrix}+\sin(\omega_{k}t)\,\begin{pmatrix}0 & \omega_{k}^{-1}\\
-\omega_{k} & 0
\end{pmatrix}\right].\nonumber 
\end{eqnarray}

In case we insist on using the original response $R$ for all later
times, the change in the uncertainty variance $D^{*}$ would have
been needed to be captured by either $\Phi'$ or by $N'$. Neither
is optimal for this, which is why the resulting schemes would lose
information in the course of the simulation. As we will see in Sect.
\ref{sub:Information-field-theoretical}, our scheme with evolving
response is lossless with respect to information.

For the data update from $d=d_{t}$ to $d'=d_{t'}$ at $t'=t+\delta t$
we need only to expand \eqref{eq:d' eq} to first order in $\delta t$.
In our ideal case with $N\rightarrow0$ we have $W'=\Phi R_{t'}^{\dagger}(R_{t'}\Phi R_{t'}^{\dagger})^{-1}=\Phi R_{t'}^{\dagger}(R\Phi R^{\dagger})^{-1}=\Phi R_{t'}^{\dagger}\tilde{\Phi}^{-1}=W-\Phi(R_{t}-R_{t'})^{\dagger}\tilde{\Phi}^{-1}\,\widehat{=}\, W-\delta t\,\Phi\, L^{\dagger}R_{t}^{\dagger}\tilde{\Phi}^{-1}$,
as a short calculation verifies. The data evolution is then 
\begin{eqnarray}
d' & = & \left(W'^{\dagger}D^{*-1}W'\right)^{-1}W'^{\dagger}D^{*-1}m^{*}\nonumber \\
 & \widehat{=} & \left(W'^{\dagger}D^{*-1}W'\right)^{-1}W'^{\dagger}D^{*-1}(1+\delta t\: L)\, W\, d\nonumber \\
 & \widehat{=} & \left(W'^{\dagger}D^{*-1}W'\right)^{-1}\times\nonumber \\
 &  & \left[W'^{\dagger}D^{*-1}W'+W'^{\dagger}D^{*-1}(W-W'+\delta t\: L\, W)\right]\, d\nonumber \\
 & \widehat{=} & d+\left(W'^{\dagger}D^{*-1}W'\right)^{-1}W'^{\dagger}D^{*-1}\delta t\,\times\nonumber \\
 &  & (\underbrace{\Phi\, L^{\dagger}+L\,\Phi}_{0})\, R_{t}^{\dagger}\,\tilde{\Phi}^{-1}d\nonumber \\
 & = & d,\label{eq:d'=00003Dd}
\end{eqnarray}
since $\Phi\, L^{\dagger}=\beta^{-1}E^{-1}E\, S^{\dagger}=-S\, E\,\beta^{-1}E^{-1}=-L\,\Phi.$
Thus $\partial_{t}d_{t}=0$, the data should not be changed, and the
evolution is completely captured by the response evolution. This scheme
is optimal from an IFT point of view as we will see in the following.
Note, that the scheme is completely specified in the data space, since
\eqref{eq:d'=00003Dd} does not require any sub-grid calculations
as it does not require any calculations at all. It will be shown in
the next section that the evolution of binned field values is also
completely specified in data space and that the sub-grid field configuration
predictions require the usage of a finer grid only at the very end
of the calculation. 

This simple data (non-)evolution equation $\partial_{t}d_{t}=0$ is
a consequence of our KG example having a linear symplectic evolution,
as determined by $\mathcal{H}(\phi)=\frac{1}{2}\phi^{\dagger}E\,\phi$
and a thermal prior distribution, as characterized by $H(\phi|\beta)=\beta\mathcal{H}(\phi)$,
both depending on the same energy matrix $E$. In general, $\partial_{t}d_{t}\neq0$
can be expected as soon as the prior and dynamics are more orthogonal
in their eigenvector sets. $\blacktriangleleft$

\subsection{Information field theoretical solution\label{sub:Information-field-theoretical}}

$\blacktriangleright$ The KG problem is exactly solvable and the
later time field can be obtained from applying a time translation
operator, as given by \eqref{eq:T_time-translation}, to an earlier
time field. This operator depends only on the time difference, $\phi_{t'}=T_{t'-t}\phi_{t}$,
and is even invertible, so that the earlier field can be calculated
from the later one. With this, the time invariance of the field covariance
can easily be verified, 
\begin{equation}
\Phi_{t}=\langle\phi_{t}\phi_{t}^{\dagger}\rangle_{(\phi)}=T_{t}\Phi_{t=0}T_{t}^{\dagger}=\Phi_{0}\equiv\Phi,\label{eq:Phi_T =00003D00003D00003D Phi}
\end{equation}
where the last identity requires a few lines of straightforward matrix
multiplications using \eqref{eq:Phikq} and \eqref{eq:T_time-translation}.

Since we want to infer the future field $\phi_{t}$ from the initial
data $d=d_{t=0}$, we have to specify how the initial data depends
on the future field. This backward-in-time response is simply given
by 
\begin{equation}
d=R\,\phi_{0}=\underbrace{R\, T_{-t}}_{R_{t}}\phi_{t}\equiv R_{t}\phi_{t}.
\end{equation}

Since we now have the response of the initial data $d=d_{0}$ to the
field $\phi_{t}$ as well its variance $\Phi_{t}$ at a later time,
we can simply write down the Wiener filter mean field at time $t$
that is 
\begin{equation}
m_{t}=\left\langle \phi_{t}\right\rangle _{(\phi_{t}|d)}=W_{t}d=\Phi R_{t}^{\dagger}\tilde{\Phi}^{-1}d.\label{eq:mt-WF}
\end{equation}
Here we used the identity $R_{t}\Phi R_{t}^{\dagger}=R\, T_{-t}\Phi T_{-t}^{\dagger}R^{\dagger}=R\,\Phi\, R^{\dagger}=\tilde{\Phi}$
that follows from \eqref{eq:Phi_T =00003D00003D00003D Phi}. Therefore,
any future mean field can be calculated directly from the original
data, which therefore does not need to be evolved in time. The response
$R_{t}$ and Wiener filter $W_{t}$ operators connecting the field
at time $t$ to the static data $d=d_{t=0}$ are exactly the ones
which were found for the ideal IFD scheme. Thus, IFD reproduces IFT
if the parameters of the future instances are able to capture all
details of the evolved PDF.%
\footnote{The observation that an entropic matching approximation enforced in
any instance of continuous time can result in the exact equation for
a dynamical system was observed previously in an attempt to reconstruct
quantum mechanics from statistics \citep{2009AIPC.1193...48C}. %
} The sub-grid representation of the evolved field as given by \eqref{eq:mt-WF}
only requires complex operations in data space, since $\tilde{\Phi}$
is fully specified there. Solely the back-projection into continuous
signal space by $R_{t}^{\dagger}$ and the subsequent spectral weighting
by $\tilde{\Phi}$ require sub-grid operations. 

One might therefore ask how the virtual data $\tilde{d}_{t}=R\,\phi_{t}$
of the original response $R$ applied to later field configurations
would evolve and if this requires a sub-grid field resolution. This
is of importance to us, since we want to compare the IFD/IFT scheme
with ad-hoc schemes, which do not need to have a notion of a sub-grid
structure. Since the future field is not precisely known, the correct
data at later times can not be specified. The best we can do is to
calculate the a posteriori expectation value of this hypothetical
future data. This ideal data at later time, $\check{d_{t}}\equiv\langle\tilde{d}_{t}\rangle=\left\langle R\phi_{t}\right\rangle _{(\phi_{t}|d)}$,
is therefore 
\begin{equation}
\check{d_{t}}=\underbrace{R\Phi R_{t}^{\dagger}(R\Phi R^{\dagger})^{-1}}_{\tilde{T}_{t}}d\equiv\tilde{T}_{t}d.\label{eq:dtilde}
\end{equation}

Note that the time translation operator of the data $\tilde{T}_{t}$
is not unity in general, basically it is only $\tilde{T_{t}}=\mathbb{1}$
for $t=0$, since one of the response operators contains a time translation
of the field: 
\begin{eqnarray}
\left(\tilde{T}_{t}\right)_{kq} & = & \left(R\Phi T_{-t}^{\dagger}R^{\dagger}\tilde{\Phi}^{-1}\right)_{kq}\label{eq:Ttilde}\\
 & = & \sum_{k'\in k+\mathcal{N}\mathbb{Z}}\frac{2\,(1-\cos(k'\Delta))}{k'^{2}\Delta^{2}}\,\times\nonumber \\
 &  & \begin{pmatrix}\begin{array}{lr}
\omega_{k'}^{-2}\cos(\omega_{k'}t) & \quad-\omega_{k'}^{-1}\sin(\omega_{k'}t)\\
\omega_{k'}^{-1}\sin(\omega_{k'}t) & \cos(\omega_{k'}t)
\end{array}\end{pmatrix}\tilde{\Phi}_{kq}^{-1}.\nonumber 
\end{eqnarray}

\begin{figure*}[t]
\includegraphics[width=0.5\textwidth]{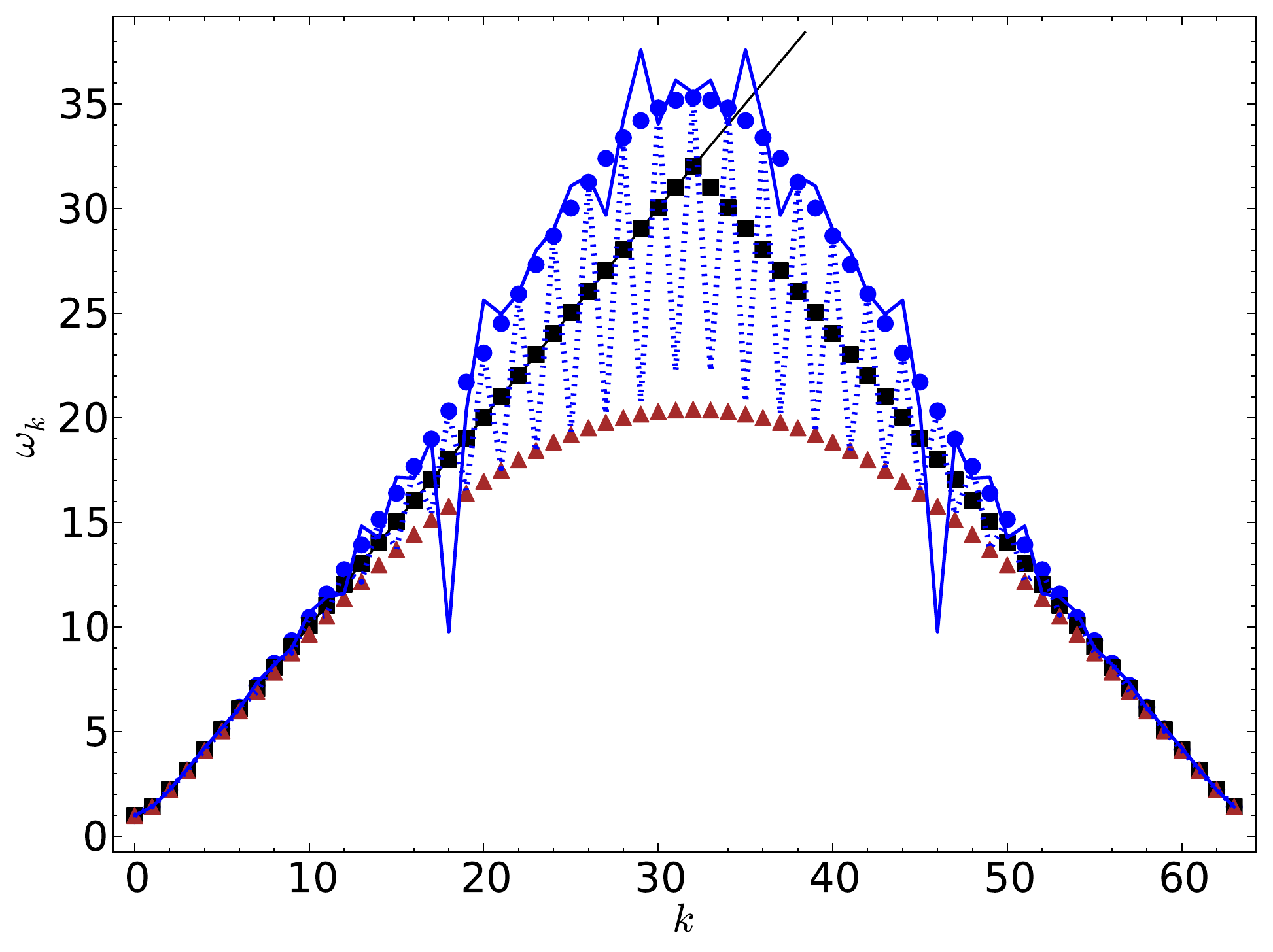}\hspace{-0.5\textwidth}{\small (a)}\hspace{-1.5em}\hspace{0.5\textwidth}\includegraphics[width=0.5\textwidth]{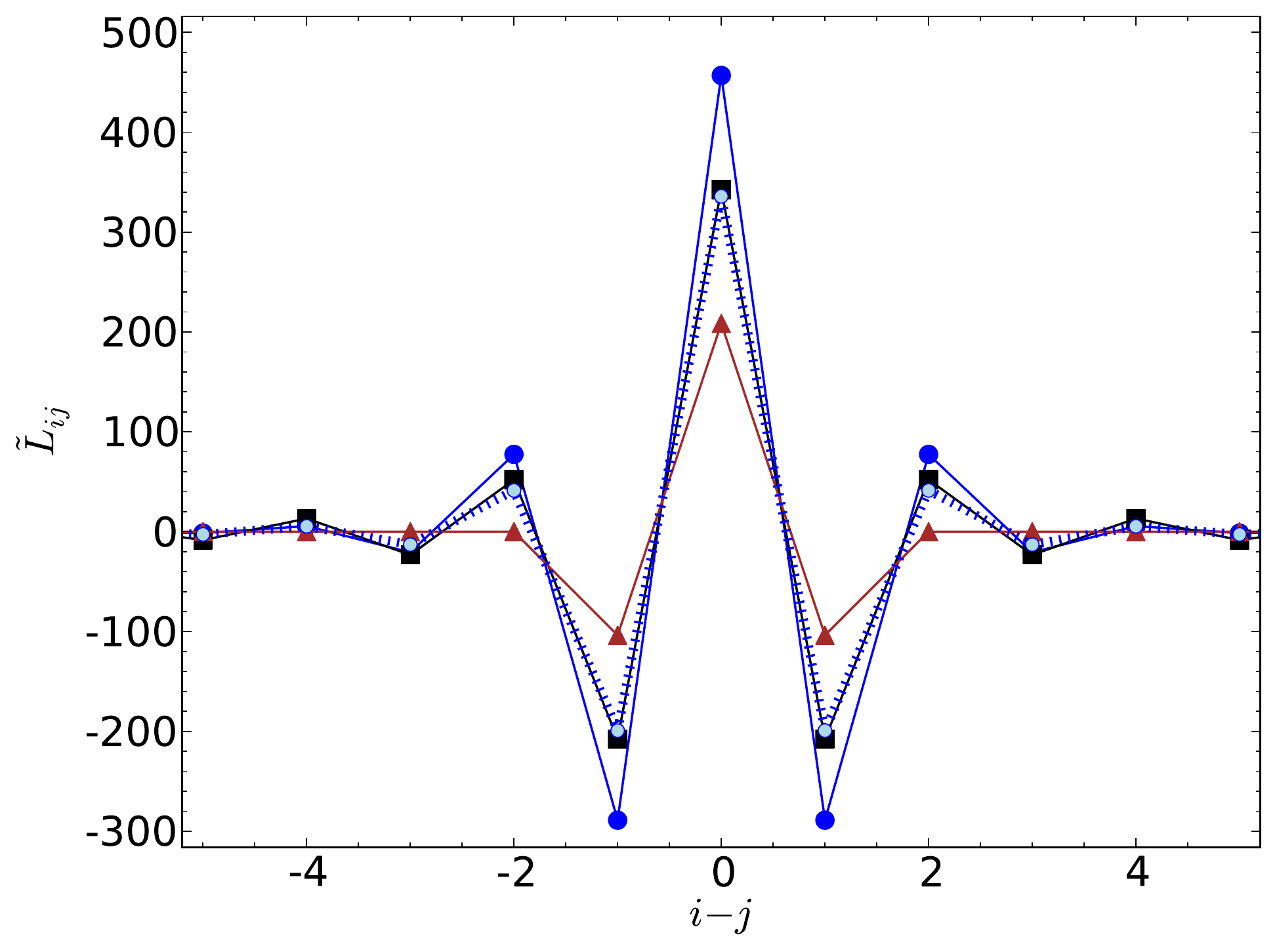}\hspace{-0.5\textwidth}{\small{}
(b)}\hspace{-1.5em}\hspace{0.5\textwidth}

\caption{(Color online) \label{fig:Fourier-data-space-dispersion}\textbf{(a):}
Fourier-data space dispersion relations $\tilde{\omega}_{k}$ of numerical
schemes for the KG field simulation for the parameters $\mathcal{N}=64$
and $\mu=1$. The IFD scheme data mode frequencies $\check{\omega}_{k,t}$
are shown at initial time $t=0$ as given by \eqref{eq:omega-data}
(top, blue dots), an instance later at $t=10^{-4}$ (top, blue solid
line with kinks), and at time $t=\pi/2$ (strongly oscillating blue
dotted line). At $t=\pi,$ the IFD scheme dispersion relation looks
similar to the initial one. The spectral scheme frequencies $\tilde{\omega}_{k}^{\mathrm{spec}}$
as given by \eqref{eq:omega2-spec} (middle, black squares) follow
the continuous space field-dispersion (thin, smooth, and black line).
Finally, the finite difference scheme $\tilde{\omega}_{k}^{\mathrm{diff}}$
as given by \eqref{eq:omega2diff} has the lowest frequencies (bottom,
brick red triangles). \textbf{(b):} Data space representation of the
numerical scheme operator $\tilde{L}_{i\, j}$ as a function of the
pixel number difference $i-j$ for small differences. The curves are
given by the discrete Fourier transformations of $\tilde{\omega}_{k}^{2}$
for the IFD scheme at $t=0$ (most extreme, blue dots and line) as
well as for $t=\pi/2$ (smaller light blue dots and blue dotted line
close to intermediate black line), the spectral scheme (intermediate
values, black squares and line), and for the finite difference scheme
(most moderate values, brick red triangles and line). It should be
noted that the IFD operator at $t=\pi/2$ also contains some power
around positions $i-j=\pm\mathcal{N}/2=\pm32$ (not shown in this
figure) as a consequence of the heavy oscillations of $\check{\omega}_{k,t}$
at this time that are visible in panel (a).}
\end{figure*}

Since this time evolution operator is fully determined in data space,
and the sub-grid mode dynamics is just captured by a sum in a pre-factor
to the computational expensive operator $\tilde{\Phi}_{kq}^{-1}$,
we can conclude that a data space only scheme was derived. The time
evolving data $\check{d_{t}}=\langle\tilde{d}_{t}\rangle$ contains
the same information as $d$, since the latter can be reconstructed
from the former via $d=\tilde{T}_{t}^{-1}\check{d}_{t}.$ We can derive
an evolution equation for $\check{d}_{t}$ by simply taking the temporal
derivative of \eqref{eq:dtilde}: 
\[
\partial_{t}\check{d}_{t}=(\partial_{t}\tilde{T}_{t})d=(\partial_{t}\tilde{T}_{t})\,\tilde{T}_{t}^{-1}\check{d}_{t}.
\]
It is obvious that this ideal evolution equation of the virtual data
according to the original response $R$ is not only more complicated
than just having an evolving response $R_{t}$ and stationary data,
it is also a differential equation with time dependent coefficients.
This might be surprising, since the dynamical equation of the underlying
KG field is invariant under time translation. However, this time-translational
symmetry is broken for our knowledge state on the field, for which
the time $t=0$ of the initial data set $d=R\phi_{t=0}$ is clearly
singled out. The different Fourier data modes are mixtures of different
field modes, which evolve with individual frequencies. Thus, the recovery
of a similar mixture $\tilde{d}_{k}=(R\,\phi_{t})_{k}=\sum_{i\in\mathbb{Z}}2\,\pi\, e^{-\frac{1}{2}\iota k\Delta}\mathrm{sinc}\left(\frac{1}{2}k\Delta+\pi\, i\right)\,(T_{t}\phi)_{k+\mathcal{N}i},$
with the original phases in the response works differently at different
times, due to the changed phases of the individual modes. Therefore,
the optimal IFD differential equation for data according to the original
response becomes time dependent. Nevertheless, we would like to have
something like a (now time dependent) data mode frequency for a comparison
with ad-hoc simulation schemes. An observer of the data dynamics could
estimate such a frequency in a pragmatic way by using $\partial_{t}^{2}\check{d_{k}}+\check{\omega}_{k,t}^{2}\check{d_{k}}=0$
as an analog of $\partial_{t}^{2}\varphi_{k}+\omega_{k}^{2}\varphi_{k}=0$
to define 
\begin{equation}
\check{\omega}_{k,t}^{2}=-(\partial_{t}^{2}d_{k,t}^{(\varphi)})/d_{k,t}^{(\varphi)}.
\end{equation}
The resulting frequencies are best calculated numerically, since the
involved formula \eqref{eq:Ttilde} contains an infinite sum without
a known closed forms. For $t=0$, however, a closed form can be derived,
\begin{eqnarray}
\check{\omega}_{k,t=0}^{2} & = & \mu^{2}\left(1-\frac{2}{\Delta\mu}\:\frac{\sinh\left(\mu\Delta\right)\,\sin\left(\frac{1}{2}\, k\Delta\right)^{2}}{\cosh\left(\mu\Delta\right)-\cos\left(k\Delta\right)}\right)^{-1}\label{eq:omega-data}\\
 & = & (k^{2}+\mu^{2})\,(1+\frac{k^{2}\Delta^{2}}{12}+\mathcal{O}(\Delta^{4})),\nonumber 
\end{eqnarray}
that recovers the original continuous space KG frequency $\omega_{k}=(k^{2}+\mu^{2})^{\nicefrac{1}{2}}$
in the limit $\Delta\rightarrow0$, but differs from it for finite
grid spacings. The oscillation frequency of a data mode is slightly
higher than the directly corresponding continuous field mode, since
the former also contains field modes from larger $k$, which have
larger frequencies, due to the mode mixing of the response operator.
The advanced revolution of the field modes at early times will be
compensated later on by a reduced oscillation speed. The initial and
later time data dispersion relation is shown in Fig. \ref{fig:Fourier-data-space-dispersion}
together with those of ad-hoc schemes derived in the next section.
$\blacktriangleleft$

\subsection{Summary of the derivation}

$\blacktriangleright$ A brief summary of the essential steps of the
IFD recipe applied to the KG problem might be instructive: 
\begin{enumerate}
\item \textbf{Field dynamics:} The KG equation was converted into a differential
equation of first order in time, $\partial_{t}\phi=L\,\phi$, by the
introduction of the momentum field $\pi_{x}=\dot{\varphi}_{x}$ as
a second component of a two component field $\phi=(\varphi,\pi)^{\dagger}$.
The KG equation is a linear as $L$ is independent of $\phi$. This
simplified the derivation of an IFD scheme. If a nonlinear equation
should be simulated, the equation has to be linearized around the
current mean field at any simulation time step.
\item \textbf{Prior knowledge:} The a priori KG field statistics was specified
as a thermal distribution $\mathcal{P}(\phi)\propto\exp(-\mathcal{H}(\phi)/T)$.
The fact that in this case the KG Hamiltonian $\mathcal{H}(\phi)$
determines both the dynamical operator $L$ as well as the a priori
statistics $\mathcal{P}(\phi)$ turns out to simplify the resulting
scheme considerably. It is, however, not a general necessity for the
applicability of IFD. The a priori distribution is a Gaussian since
the Hamiltonian is quadratically in $\phi$. If non-Gaussian priors
are to be used it is recommended to find a Gaussian approximation
since IFD is developed so far only for Gaussian priors. 
\item \textbf{Data constraints:} As a next step, the computer data space
was introduced. The computer data $d$ needs to be related to the
field $\phi$ and this relation should be linear for practical reasons
and could be assumed to be noiseless for the KG example, $d=R\,\phi$.
The initial discretization operator $R$ was chosen here to perform
a simple bin average. Therefore the average field value in each bin
is known if the data is available, but not the detailed field configuration
within those. However, not all possible sub-grid field configurations
are equally plausible, since the prior gives them different weights.
Combining prior and data information, the ensemble of plausible field
configurations can be specified, and characterized by its mean field
$m=\langle\phi\rangle_{(\phi|d)}$ and uncertainty variance $D=\langle(\phi-m)\,(\phi-m)^{\dagger}\rangle_{(\phi|d)}$
determining a Gaussian a posteriori distribution $\mathcal{\mathcal{P}}(\phi|d)=\mathcal{G}(\phi-m,\, D)$.
This is a Gaussian thanks to the Gaussian prior and linear data model.
The mean field and its variance are auxiliary mathematical objects
used in the derivation of the simulation scheme that need not concrete
representations in computer memory.
\item \textbf{Field evolution:} The action of the time evolution operator
on the posterior distribution had then to be worked out analytically.
Since we insisted on linear or linearized operators, the time evolved
posterior is again a Gaussian, $\mathcal{\mathcal{P}}(\phi'|d)=\mathcal{G}(\phi'-m^{*},\, D^{*})$,
characterized by an updated mean $m^{*}$ and uncertainty variance
$D^{*}$, both again auxiliary mathematical objects.
\item \textbf{Prior update:} The prior of the later time might be different
and should be updated since it will be used again. However, due to
energy and phase space conservation of the KG dynamics, the KG prior
is unchanged. This step could have been skipped, since the evolution
of the prior can also be determined as part of the next step, the
data update via entropic matching. However, this requires that the
field dynamical equation captures all sub-grid physics. If this not
the case, the prior update step might permit to implement sub-grid
processes not being present in the dynamical equation.
\item \textbf{Data update:} Finally, an update formula for the later time
data $d'$ in computer memory was constructed. This was done by first
specifying the mathematical relationship between any such data and
the later time field a posteriori distribution, $\mathcal{\mathcal{P}}(\phi'|d')=\mathcal{G}(\phi'-m',\, D')$,
where $m'=D'\, R'^{\dagger}d'$ and $R'$ and $D'$ are response and
propagator/variance at the later time. Then the time evolved distribution
$\mathcal{\mathcal{P}}(\phi'|d)$ and the one determined by the new
data $\mathcal{\mathcal{P}}(\phi'|d')$ were matched entropically.
The parameters used to get an optimal match can be any of the later
time, primed quantities. In the particular KG example it turned out
to be most effective to vary $d'$ and $R'$ in the entropic matching
since this way an information-lossless scheme could be obtained. This
scheme maps the entire field evolution onto an evolving response operator
$R_{t}$ and stationary data. We showed that the resulting simulation
scheme is indeed optimal by comparison to the exact information theoretically
derived solution of the future field prediction problem. Since this
particular KG simulation scheme does not modify the data, we asked
how the binned field values (with stationary bin-averaging) would
evolve and derived their evolution equation. The time translation
operator of this does also not require any explicit sub-grid field
representation, but has encodes sub-grid physics implicitly. 
\end{enumerate}
The derived simulation scheme can now be implemented on a computer.
The resulting code performs only data space operations and does not
require any sub-grid representation. The sub-grid physics, the prior
knowledge, and the details of the measurement process (the data to
fields relation) have all been included in the IFD scheme.

\section{Numerical verification\label{sec:Numerical-verification}}

\subsection{Standard simulation schemes}

$\blacktriangleright$ The IFD scheme for the KG field should now
be compared to more standard simulation schemes for the KG equation
as described in Appendix \ref{sub:Discretization-of-differential}.

The most common one is the finite difference discretization of the
differential operators by setting $\partial_{x}\varphi_{x}\approx(\varphi_{(i+1)\Delta}-\varphi_{i\Delta})/\Delta$
and $\partial_{x}^{2}\varphi_{x}\approx(-\varphi_{(i+1)\Delta}+2\varphi_{i\Delta}-\varphi_{(i-1)\Delta})/\Delta^{2}$.
The KG equation discretized in this way, $\partial_{t}d=\tilde{L}^{\mathrm{diff}}\, d$
with $\tilde{L}_{ij}^{\mathrm{diff}}=\Delta^{-2}\delta_{i\,[j+1]_{\mathcal{N}}}-(2\Delta^{-2}+\mu^{2})\,\delta_{ij}+\Delta^{-2}\delta_{i\,[j-1]_{\mathcal{N}}}$
and $[j]_{\mathcal{N}}=j\,\mathrm{mod}\,\mathcal{N}$, becomes diagonal
in Fourier space, just with the dispersion relation given by 
\begin{equation}
\omega_{k}^{2}\rightarrow\left(\tilde{\omega}_{k}^{\mathrm{diff}}\right)^{2}=\mu^{2}+2\Delta^{-2}(1-\cos\left(k\Delta\right)).\label{eq:omega2diff}
\end{equation}

This and the IFD dispersion relation are shown in Fig. \ref{fig:Fourier-data-space-dispersion}
in comparison to the one of the original KG field, $\omega^{2}=\mu^{2}+k^{2}.$
Since the initial IFD frequencies are above, and the frequencies of
the difference scheme are below the one of the KG field, it is also
natural to consider the latter as another option. Thus we also investigate
a spectral simulation scheme with:%
\footnote{The distinctions of the cases is only necessary here, since we use
$k\in\{0,\,\ldots\,\mathcal{N}-1\}$ so that the negative frequencies
are represented by wave numbers in the second half of the range. If
we would use $k\in\{-\mathcal{N}/2+1,\,\ldots\,\mathcal{N}/2\}$ as
our first Brillouin zone, we would have $\left(\tilde{\omega}_{k}^{\mathrm{spec}}\right)^{2}=\mu^{2}+k^{2}$.%
} 
\begin{equation}
\left(\tilde{\omega}_{k}^{\mathrm{spec}}\right)^{2}=\begin{cases}
\mu^{2}+k^{2} & \;\mbox{for}\: k\in\{0,\,\ldots\,\mathcal{N}/2\}\\
\mu^{2}+(\mathcal{N}-k)^{2} & \;\mbox{for}\: k\in\{\mathcal{N}/2,\,\ldots\,\mathcal{N}\}
\end{cases}.\label{eq:omega2-spec}
\end{equation}

The Fourier space data evolution equation can be solved analytically
and has the solution
\begin{eqnarray}
d_{k}^{(\varphi)} & = & \tilde{a}_{k}e^{\iota\tilde{\omega}_{k}t}+\overline{\tilde{a}_{\mathcal{N}-k}}e^{-\iota\tilde{\omega}_{k}t}\nonumber \\
d_{k}^{(\pi)} & = & \iota\tilde{\omega}_{k}\left(\tilde{a}_{k}e^{\iota\tilde{\omega}_{k}t}-\overline{\tilde{a}_{\mathcal{N}-k}}e^{-\iota\tilde{\omega}_{k}t}\right),\label{eq:data-evolution-solution}
\end{eqnarray}
with the coefficients determined by the initial data 
\begin{equation}
\tilde{a}_{k}=\frac{d_{k\, t=0}^{(\varphi)}}{2}+\frac{d_{k\, t=0}^{(\pi)}}{2\iota\tilde{\omega}_{k}}.\label{eq:initial-conditions}
\end{equation}
Thus, the most efficient simulation scheme for the KG field evolution
schemes is to evolve the initial data according to these Fourier space
equations analytically and transform the field back to position space
at the desired time.

The ad hoc simulation schemes are best implemented via \eqref{eq:data-evolution-solution}
and \eqref{eq:initial-conditions}, the corresponding data $\check{d}$
of the IFD scheme according to \eqref{eq:dtilde} and \eqref{eq:Ttilde},
whereas the full field including the sub-grid modes can be followed
via \eqref{eq:field-evolution-solution}. $\blacktriangleleft$

\begin{figure*}[t]
\includegraphics[width=0.5\textwidth]{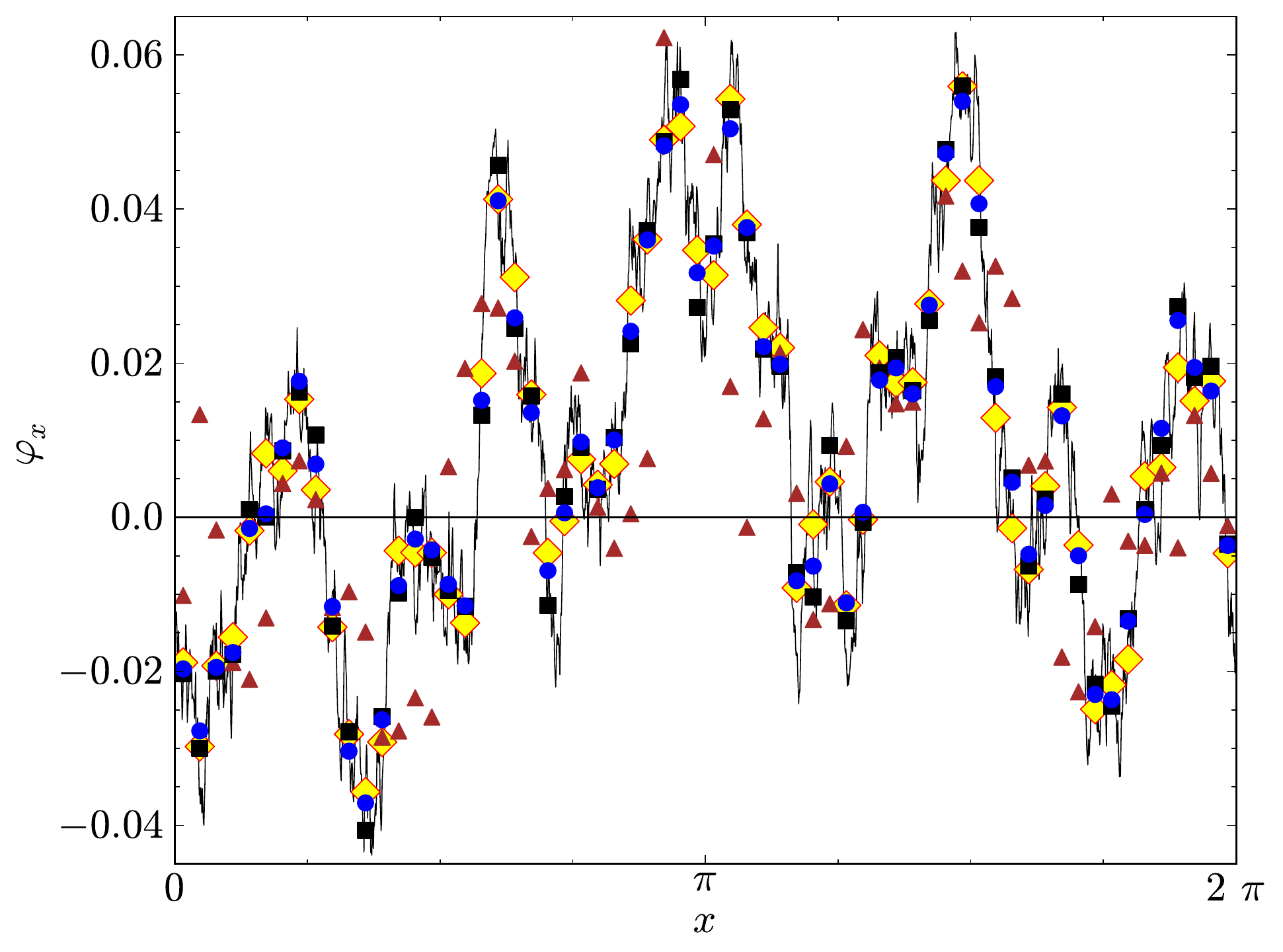}\hspace{-0.5\textwidth}{\small (a)}\hspace{-1.5em}\hspace{0.5\textwidth}\includegraphics[bb=40bp 20bp 530bp 400bp,clip,width=0.487\textwidth]{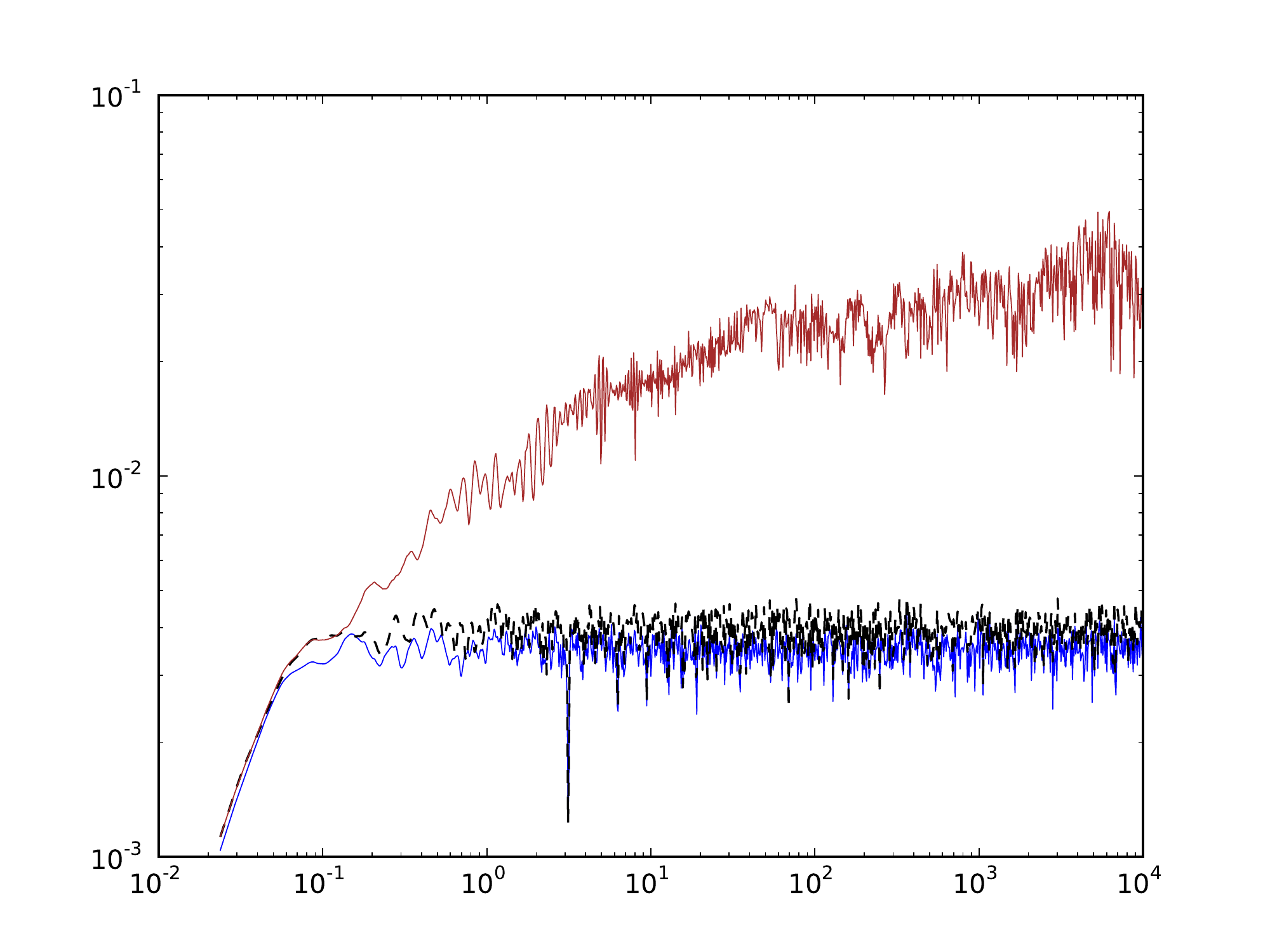}\hspace{-0.5\textwidth}{\small{}
(b)}\hspace{-1.5em}\hspace{0.5\textwidth}

\caption{(Color online)\textbf{ (a):} Evolved field (thin, black line) and
data at $t=10$ of the field also shown in Fig. \ref{fig:thermal-KG_field}
($\beta=1,\,\mu=1,\:\mathcal{N}=64$). The exact data $\tilde{d}_{t}=R\varphi_{t}$
are shown as yellow diamonds. The IFD data according to \eqref{eq:dtilde}
and \eqref{eq:Ttilde} (blue dots) follows the exact data closely.
The data of the spectral scheme (black squares) is very close to the
IFD data. The data of the difference scheme (brick red triangles)
exhibit the poorest match to the correct data of the evolved field.
The root mean square errors of the field data values $\sigma_{d}^{(\varphi)}=\sqrt{\sum_{i=0}^{\mathcal{N}-1}(\tilde{d}^{(\varphi)}-R\varphi)_{i}^{2}/\mathcal{N}}$
of the three schemes are $0.003$, $0.004$, and $0.020$ for the
IFD, spectral, and difference scheme, respectively. \textbf{(b):}
Temporal evolution of the data error $\sigma_{d}^{(\varphi)}(t)$
for the IFD (bottom solid blue line), spectral (dashed black line
slightly above the former), and finite difference (top brick red line)
scheme. The dip in the IFD and spectral scheme error at $t=\pi$ is
due to the nearly perfect alignment of the mode phases at this particular
time. \label{fig:Evolved-field}}
\end{figure*}

\subsection{Time evolution}

$\blacktriangleright$ To see how well the different simulation schemes
perform, we simulate a KG field by setting up its Fourier amplitudes
$a_{k}\in\mathbb{C}$ up to $|k|=\mathcal{N}_{\phi}/2$ drawn from
$\p(a_{k})=\mathcal{G}(a_{k},1/(4\,\beta\,(\mu^{2}+k^{2})))$ and
$a_{\mathcal{N}_{\phi}-k}=\overline{a_{k}}$ for the ``negative''
modes, so that \eqref{eq:H(a)}, \eqref{eq:thermal prior} and $\phi_{x}\in\mathbb{R}^{2}$
are satisfied. We use $\mathcal{N}_{\phi}=2048$, $\mu=1$, and $\beta=1$.
A resulting field realization is displayed in Fig. \ref{fig:thermal-KG_field}.
We time-evolve all its Fourier modes according to \eqref{eq:field-evolution-solution}.
The initial and late time exact data is generated via $\tilde{d}_{t}=R\phi_{t}$
with the response given by \eqref{eq:Rix} for $\mathcal{N}=64$ data
bins. This means that there are $\mathcal{N}_{\phi}/\mathcal{N}=32$
independent field modes combined in a single datum, ensuring that
there is substantial sub-grid uncertainty, as is well observable in
Fig. \ref{fig:thermal-KG_field}. For the spectral and difference
schemes, the data is time evolved according to \eqref{eq:data-evolution-solution}
and \eqref{eq:initial-conditions}. For the IFD scheme, we use \eqref{eq:dtilde}
and \eqref{eq:Ttilde} to calculate corresponding late time data.

For time $t=10$, the field is shown and the different data sets at
this time are compared in Fig. \eqref{fig:Evolved-field}. This time
was chosen for that the difference scheme already exhibits some significant
but still moderate deviations from the correct solution. The IFD and
spectral scheme are both relatively accurate. A difference between
them exists, but is hard to see by eye in this snapshot. However,
a comparison of the spatially averaged errors of the two schemes reveals
a significantly higher accuracy of the IFD scheme with respect to
the spectral scheme at basically all times.

Although the IFD scheme has the highest fidelity, the spectral scheme
is also very good for arbitrarily large times. The reason can easily
be understood. Despite the fact that any data Fourier mode is a mixture
of several field modes, the spectral scheme just follows the most
dominant of these modes, and treats the others as random noise. However,
since the main mode is correctly captured, it can be followed for
infinitely large intervals, and the ignored modes just contribute
a fixed amount of uncertainty. The IFD scheme also assigns some power
to these higher modes and follows their evolution. This is why it
has a higher accuracy.

Optimally, one would have chosen an initial response that maps the
first $\mathcal{N}$ Fourier modes of the field exactly into the data.
Then these modes could have been followed with absolute precision,
while one would have no information on the lower amplitude higher
Fourier modes. In this case the IFD scheme would have been identical
to the spectral scheme, but it would not have served us well as a
sufficiently complex example illustrating the inner workings of the
IFD framework.$\blacktriangleleft$

\section{Conclusions and outlook\label{sec:Conclusions-and-outlook}}

Information field dynamics serves as a framework to derive numerical
simulation schemes. It rests on information field theory in order
to construct continuous space field configurations out of the finite
data in computer memory. It uses the maximum entropy principle to
construct updated computer memory data so that the ensemble of time-evolved
continuous space field configurations is matched by the ensemble implied
by the updated data with minimal information loss.

The data updating operations of an IFD simulation time step, as given
by \eqref{eq:d' eq} and \eqref{eq:D' eq}, are in general complex,
and might require the usage of linear algebra solvers. However, for
numerical stability reasons, an implicit time step scheme might be
adopted for a simulation anyway, and the linear algebra operations
of the implicit and IFD schemes might be performed together.

As an illustrative example, we have derived the optimal IFD scheme
for a thermally excited Klein-Gordon field. It could be shown that
the resulting IFD scheme is identical to the one resulting from IFT.
The scheme is much more accurate than a simplistic real space discretization
of the differential operator, and it is still significantly more accurate
than a spectral scheme. In comparison to these two ad hoc schemes
with stationary evolution equations for the data, the IFD scheme exhibits
a time dependent discretization of the differential equation. This
is due to its ability to follow to some level the evolution of the
sub-grid scales without representing them explicitly in computer memory
but capturing their influence implicitly in the data update rules.

This initial work on IFD should be regarded as a proposal for how
to incorporate information theoretical considerations into the construction
of simulation schemes. IFD permits us to state and include explicitly
background knowledge on sub-grid behavior as well as external measurement
data in a way that hopefully exploits and conserves as much of the
available information as possible.

For technical reasons, one might compromise information theoretical
fidelity for reducing the numerical complexity. Also for this balance,
the information theoretical language introduced here should help to
judge the choices. Finally, the language of IFD is already what is
needed for data assimilation simulation schemes, as for example used
in weather forecasts. The next goal of this research line is to develop
IFD schemes for scientifically and technologically more relevant problems,
like turbulent hydrodynamics. This, however, is left for future work.

\subsection*{Acknowledgements}

I gratefully thank Michael Bell, Maksim Greiner, Henrik Junklewitz,
Ewald Müller, Niels Opermann, Tiago Ramalho, Martin Reinecke, Thomas
Riller, Marco Selig, Lars Winderling and two anonymous referees for
discussions, feedback, and comments. The calculations were performed
using the \texttt{Sage} \citep{sage-4.7} mathematics software.

\appendix
%dummy comment inserted by tex2lyx to ensure that this paragraph is not empty
%dummy comment inserted by tex2lyx to ensure that this paragraph is not empty

\section{Previous work\label{sec:Appendix----Previous}}

\subsection{Discretization of differential operators\label{sub:Discretization-of-differential}}

Most of the dynamical systems in physics are described by partial
differential equations. These contain differential operators acting
on the dynamical fields. With the finite representation of the fields
in computer memory, these operators need a discretized representation
as well. A number of discretization schemes have been developed, including
finite difference methods, finite volume methods, finite element methods,
spectral methods, smoothed particle hydrodynamics and others. Most
of these schemes assume a distinct sub-grid structure for the fields,
in contrast to IFD.

\textbf{Finite difference methods} \citep{Courant1928}, represent
differentials by finite differences between the field values at the
lattice grid points. These finite difference operators are exact if
the field is polynomial of the order of the operator. Thus a finite
difference gradient operator implicitly assumes the field to be piecewise
linear on sub-grid scales, a Laplace operator the field to be quadratic
and so forth. In Sect. \ref{sec:Numerical-verification} we will show
numerically that the IFD operator for the KG field evolution is superior
to the finite difference operator.

\textbf{Finite volume methods} \citep{Godunov1959} are used when
conserved quantities are simulated, such as e.g. the fluid mass in
hydrodynamics. The space is split into pixel volumes. The continuity
equations for the conserved quantities can be turned into balance
equations for the fluxes of the quantity through the boundaries of
a pixel's volume. The simplest assumption for the sub-grid field configuration
is that it is constant within the pixels, with jumps at their boundaries.
The resulting discontinuities have to be treated as separate Riemann
problems at the boundaries in hydrodynamics. A conservative IFD scheme
should also be possible, if the stored data of the scheme are the
amounts of the conserved quantity within pixel volumes, and the fluxes
between adjacent pixels.

\textbf{Finite element methods} \citep{Ritz1908,Galerkin1915} also
partition the space into sub-volumes, the 'elements'. A set of basis
functions for the field is defined, with a support covering only a
small number of the elements/pixels. The field is represented as a
linear combination of these basis function, and therefore with a tightly
parametrized sub-grid structure, e.g. being piecewise linear. The
partial differential equations are only required to be solved weakly,
in the Sobolev function space spanned by the chosen basis functions.
This turns spatial differential operators into linear systems of equations,
which then can be solved on a computer.

\textbf{Spectral methods} are also Sobolev space based, just with
the basis functions being Fourier modes. We will compare the IFD scheme
for the KG field to a spectral method and show that IFD provides a
slightly more accurate simulation.

\textbf{Smoothed particle hydrodynamics} \citep{1977MNRAS.181..375G,1977AJ.....82.1013L,2002MNRAS.333..649S}
discretizes the mass of the fluid and not the space. Smoothed particle
hydrodynamics is one example of Lagrangian methods, in which the 'grid'
follows the flow. Each mass element has a dynamically evolving position
and is thought to be distributed over some finite ball according to
a radially declining and adaptively sized kernel function determining
the sub-grid field structure. 

\textbf{Moving mesh codes} can be regarded as a compromise between
Eulerian schemes with fixed lattices and Lagrangian schemes with a
co-moving but particle based fluid discretization as smoothed particle
hydrodynamics \citep{1984PhyD...12..408W,1984JQSRT..31..473W}. {\small Moving
mesh codes }were recently improved by using Voronoi tessellation to
create flexible volume cells around the moving grid points on which
finite volume methods can be used \citep{2010MNRAS.401..791S}. Thus
also here the sub-grid field representation is of a predetermined
functional form.

In contrast to these approaches, IFD does not assumes an a priori
shape of the sub-grid field structure. It considers all possible sub-grid
configurations consistent with the constraints given by the data and
the field equations, but weights them with a priori plausibilities.
This requires knowledge on the sub-grid dynamics.

\subsection{Sub-grid scale modeling}

IFD, as proposed here, requires prior information on all modes of
the dynamical field, in order to constrain the unresolved degrees
of freedom. The necessity to use information on sub-grid scales in
simulations was already realized for hydrodynamics. For this reason,
the method of \textbf{large eddy simulations} was developed \citep{1963MWRv...91...99S,1970JFM....41..453D,1996AnRFM..28...45L}.
This resolves the largest scale of a flow by simulating a spatially
filtered (convolved) dynamics, in combination with sub-grid scale
models that try to summarize the effect of the unresolved scales on
the global dynamics \citep{1994ApJ...428..729C,1997ApJ...478..322C,2000ApJ...541L..79C,2011A&A...528A.106S}.
Usually stress tensors describe the sub-grid scales. These are actually
velocity fluctuation covariance matrices and therefore conceptually
similar to the uncertainty dynamical field covariances in IFD. Large
eddy simulations have recently been combined with adaptive mesh refinement
methods that increase the resolution at locations where small scale
dynamics is particularly important. This is especially important in
astrophysical applications, where a large range of scales should be
followed, as for example in galaxy clusters \citep{2009ApJ...707...40M,2012arXiv1202.5882V}. 

In \textbf{astrophysical hydrodynamics}, many additional processes
on unresolved scales, like star formation and radiative feedback,
are relevant yet cannot be followed in detail. In simulations of galaxies
using smooth particle hydrodynamics, the interstellar medium is often
described as a mixture of interacting gas phases (molecular, ionized,
...) forming a complex weather, with a single \textbf{effective equation
of state} summarizing these phases \citep{2003MNRAS.339..289S}. However,
the translation of sub-grid physics into a concrete simulation scheme
is usually done ad-hoc without considering the resolution dependent
level of sub-grid fluctuations. 

In \textbf{oceanography}, it has been recognized that some information
about sub-grid eddy evolution is contained in the large scale fluid
motions due to the practical incompressibility of water and the resulting
solenoidality of the flow patterns. Partial \textbf{reconstruction
of the sub-grid eddies} from a coarse resolution is therefore possible
\citep{1984JAtS...41.1881B}. This has been used to construct accurate
simulation schemes for advective tracers and for vorticity transport
\citep{2003cond.mat..5205B,LeSommer2011154}. A \textbf{maximum entropy
production principle} was introduced in this context in order to construct
sub-grid configurations that are numerically stable \citep{2003cond.mat..5205B}.
There, maximum entropy was regarded merely as a numerical regularization
trick, while in our work, it plays an important role in ensuring optimal
information flow between the simulation data at different time steps.

\subsection{Data assimilation methods\label{sub:Data-assimilation-methods}}

Data assimilation methods are probably most similar in spirit to IFD.
Data assimilation methods are used in weather forecast calculations
to impose constraints from past measurements on numerical simulation
of the atmosphere. A recent comparison of such methods can be found
in \citep{2011arXiv1107.4118L}. The gold standard of the field is
the full Bayesian posterior distribution of the dynamical system given
all data. Typically, there are two broad classes of algorithms used
to approximate this in a computationally affordable way: particle
ensemble filters and variational methods.

\textbf{Particle filter }represent the knowledge and uncertainty on
the system state as an ensemble of realizations, called the particles.
These evolve individually according to the system dynamics to later
times, when new measurements are available. Then, the particles are
selected and/or re-weighted according to their individual consistency
with the new data. Resampling this distribution with a new set of
particles (now with equal weights) closes the loop and prepares for
the next simulation time step. A recent discussion of such methods
can be found in \citep{2010QJRMS.136.1991V}.

\textbf{Ensemble Kalman filters} represent the system knowledge as
well as an ensemble of realizations that can be propagated by the
full non-linear dynamics in time. The data assimilation step, however,
is not done via re-weighting or re-sampling, but by Kalman filtering.
Kalman filtering is basically Wiener filtering, which we introduce
in Sect. \ref{sub:Information-field-theory}, while using an empirically
determined signal covariance matrix. This is computed from the ensemble,
which is informed by the actual external measurement data. 

\textbf{Variational methods for data assimilation} combine the action
of a Lagrangian determining the dynamics and a loss function describing
a penalty for any mismatch of the model prediction and the data \citep{bennett2002inverse}.
From this combined Lagrangian, combining dynamics and data constraints,
a variational equation aries that satisfies both the system dynamics
and the data constraints. Variational methods treat information processing
and field dynamics simultaneously, similar to IFD.

A third approach to data assimilation has recently been proposed for
the simulation of cosmic structure formation \citep{2012arXiv1203.3639J,2012arXiv1203.4184K,2010MNRAS.403..589K}.
There the full posteriori of the cosmic matter field as determined
by galaxy catalogs and the Gaussian initial condition statistics of
cosmic structure formation is sampled via a \textbf{Hamiltonian sampling}
method.

\section{Maximum entropy principle\label{sec:Appendix----Maximum}}

The MEP \citep{1957PhRv..108..171J,1957PhRv..106..620J,1982ieee...70..939J,2008arXiv0808.0012C}
is uniquely specified by the following three requirement on how probabilities
should be ranked and updated with respect to new information. Entropy
is defined to quantify how well a given PDF represents a knowledge
state. Its functional form is determined by three requirements on
the resulting probability updating scheme: 
\begin{itemize}
\item \textbf{Locality:} Local information has local effects; information
that affects only some part of the phase space should not modify the
entropy and the implied MEP PDF in case this area is discarded. 
\item \textbf{Coordinate invariance:} The system of coordinates of the phase
space does not carry information. Entropy should be invariant under
coordinate transformation as well as the determined MEP PDF. 
\item \textbf{Independence:} Independent systems can be treated jointly
or separately, yielding the same entropy in both cases. The joint
MEP PDF must therefore be separable into a product of PDFs for the
individual systems. 
\end{itemize}
The unique (up to trivial rescaling) entropy functional on PDFs that
is consistent with these requirements is given by \eqref{eq:entropy}
as it was shown by \citep{1957PhRv..108..171J,1957PhRv..106..620J,1982ieee...70..939J,2008arXiv0808.0012C}.
The usual way to use this entropy in order to specify the PDF $\p(\phi)$
is to maximize it subject to some constraints imposed on certain moments
of the signal field statistics. An obvious one is the proper normalization
$\left\langle 1\right\rangle _{\p(\phi)}=1$ of the PDF, but also
a number of higher moments might be known a priori, and summarized
in the form $\left\langle f_{i}(\phi)\right\rangle _{\p(\phi)}=a_{i}$.
Here the functions could be simple moments like $\phi$, $\phi\phi^{\dagger},$
etc. or more complicated functions thereof. These constraints on PDF
moments are then incorporated into the entropy via Lagrange multiplier
or thermodynamical potentials $\mu$ and $\lambda=(\lambda_{i})_{i}$:

\begin{eqnarray}
 &  & \mathcal{S}(\p,\mu,\lambda|\mathcal{Q})\label{eq:entropy-2}\\
 & = & \mathcal{S}(\p|\mathcal{Q})-\left\langle \mu+\lambda^{\dagger}f(\phi)\right\rangle \nonumber \\
 & = & -\int\mathcal{D}\phi\:\p(\phi)\:\left[\log\left(\frac{\p(\phi)}{\mathcal{Q}(\phi)}\right)+\mu+\lambda^{\dagger}f(\phi)\right].\nonumber 
\end{eqnarray}
Maximizing this entropy with respect to all components of $\p(\phi)$
yields 
\begin{equation}
\p(\phi)=\frac{\mathcal{Q}(\phi)}{Z(\lambda)}\, e^{-\lambda^{\dagger}f(\phi)},
\end{equation}
where 
\begin{equation}
Z(\lambda)=\int\mathcal{D}\phi\, Q(\phi)\, e^{-\lambda^{\dagger}f(\phi)}
\end{equation}
ensures proper normalization, and theLagrange potentials $\lambda$
have to be chosen to satisfy 
\begin{equation}
-\partial_{\lambda}\mathcal{S}=\partial_{\lambda}\log Z=\int\mathcal{D}\phi\:\p(\phi)\: f(\phi)=\left\langle f(\phi)\right\rangle _{\p(\phi)}=a.
\end{equation}

In Sect. \ref{sub:Prior-knowledge}, it is claimed that the MEP distribution
for $\phi$ with known mean $\psi$ and covariance $\Phi$ is the
Gaussian $\mathcal{G}(\phi-\psi,\Phi).$ This can now be verified
by a short calculation. The entropy \eqref{eq:entropy-2} can be constrained
by the knowledge of zero, first, and second moments of the field via
the Lagrange-multiplier scalar $\mu$, field $\lambda$, and matrix
$\Lambda$, respectively:

\begin{eqnarray}
 &  & \mathcal{S}(\p,\mu,\lambda,\Lambda|\mathcal{Q})\\
 & = & \mathcal{S}(\p|\mathcal{Q})-\mu-\lambda^{\dagger}\left\langle \phi\right\rangle _{(\phi)}-\mathrm{Tr\left(\Lambda\,\left\langle \phi\phi^{\dagger}\right\rangle _{(\phi)}\right)}\nonumber \\
 & = & -\int\mathcal{D}\phi\:\p(\phi)\:\left[\log\left(\frac{\p(\phi)}{\mathcal{Q}(\phi)}\right)+\mu+\lambda^{\dagger}\phi+\phi^{\dagger}\Lambda\,\phi\right].\nonumber 
\end{eqnarray}
Minimizing this with respect to all components of $\p(\phi)$ for
a flat prior-prior $\mathcal{Q}(\phi)=\mathrm{const}$ subject to
the constraints 
\begin{eqnarray}
-\partial_{\mu}\mathcal{S} & =\left\langle 1\right\rangle _{(\phi)}= & 1,\\
-\partial_{\lambda}\mathcal{S} & =\left\langle \phi\right\rangle _{(\phi)}= & \psi,\\
-\partial_{\Lambda}\mathcal{S} & =\left\langle \phi\,\phi^{\dagger}\right\rangle _{(\phi)}= & \Phi+\psi\,\psi^{\dagger},
\end{eqnarray}
to ensure proper PDF normalization, mean, and variance, respectively,
yields $\p(\phi|\psi,\Phi)=\mathcal{G}(\phi-\psi,\Phi)$ as assumed
in \eqref{eq:Gauss-with-mean}.

\bibliography{../Bib/ift}

\end{document}